\documentclass[a4paper, 12pt]{article}
\usepackage[margin = 1in]{geometry}
\usepackage{amsfonts, amsmath, amssymb}
\usepackage{hyphenat}
\usepackage{fancyhdr}
\usepackage[nottoc, notlot, notlof]{tocbibind}
\usepackage{graphicx}
\usepackage{float}
\usepackage{upgreek}
\usepackage{mathrsfs,amsmath} 
\usepackage{mathtools}
\usepackage{color}
\usepackage{perpage}
\usepackage{tikz}
\usepackage{relsize}
\usepackage{siunitx}
\usepackage[hyphens,spaces,obeyspaces]{url}
\usepackage{pgfgantt}
\usepackage[section] {placeins}
\usepackage{siunitx}

\usepackage[justification=centering]{caption}

\newcommand{\etal}{\textit{et al}.}
\newcommand{\ie}{\textit{i}.\textit{e}.}
\newcommand{\angstrom}{\mbox{\normalfont\AA}}

\fancypagestyle{abstract}{%
\fancyhead[LE,LO]{}
\fancyhead[RE,RO]{}
\fancyhead[CE,CO]{APPENDIX}
\fancyfoot[CO,CE]{\thepage}
}

\DeclarePairedDelimiter\set\{\}

\MakePerPage{footnote}

\begin{document}

\begin{titlepage}
\begin{center}
\vspace*{1cm}
\huge{\textbf{Imperial College London}}\\
\huge{\textbf{Department of Physics}}
\vfill
\line(1,0){400}\\[1mm]
\Huge{\textbf{Graphene Field Effect Transistors}}\\[2mm]
\line(1,0){400}\\
\vfill
By Mohamed Warda and Khodr Badih\\
20 July 2021
\end{center}
\end{titlepage}

\begin{abstract}
The past decade has seen rapid growth in the research area of graphene and its application to novel electronics.  With Moore's law beginning to plateau, the need for post-silicon technology in industry is becoming more apparent. Moreover, existing technologies are insufficient for implementing terahertz detectors and receivers, which are required for a number of applications including medical imaging and security scanning. Graphene is considered to be a key potential candidate for replacing silicon in existing CMOS technology as well as realizing field effect transistors for terahertz detection, due to its remarkable electronic properties, with observed electronic mobilities reaching up to $\SI{2e5}{\cm\squared \per \V \per \s}$ in suspended graphene samples. This report reviews the physics and electronic properties of graphene in the context of graphene transistor implementations. Common techniques used to synthesize graphene, such as mechanical exfoliation, chemical vapor deposition, and epitaxial growth are reviewed and compared. One of the challenges associated with realizing graphene transistors is that graphene is semimetallic, with a zero bandgap, which is troublesome in the context of digital electronics applications. Thus, the report also reviews different ways of opening a bandgap in graphene by using bilayer graphene and graphene nanoribbons. The basic operation of a conventional field effect transistor is explained and key figures of merit used in the literature are extracted. Finally, a review of some examples of state-of-the-art graphene field effect transistors is presented, with particular focus on monolayer graphene, bilayer graphene, and graphene nanoribbons.
\thispagestyle{empty}
\end{abstract}

\pagebreak

\thispagestyle{empty}
\tableofcontents
\thispagestyle{empty}
\clearpage

\section*{Acronyms and Constants}
\pagenumbering{roman}

\addcontentsline{toc}{section}{Acronyms and Constants}

\textbf{\Large{Acronyms}}

\begin{table}[H]

\begin{tabular}{l l}
CMOS & Complementary metal oxide  semiconductor\\
CNT & Carbon nanotube\\
CVD & Chemical vapor deposition\\
HEMT & High electron mobility transistor\\
FET & Field effect transistor\\
GNR & Graphene nanoribbon\\
RF & Radio Frequency\\
STM & Scanning tunneling microscopy\\
\end{tabular}
\end{table}
\vspace{6mm}

\thispagestyle{plain}

\pagebreak

\textbf{\Large{Constants}}

\begin{table}[H]

\begin{tabular}{l l}

$i$ & The imaginary unit, $\sqrt{-1}$\\
$c$ & The speed of light in vacuum, $\SI{3.00e-8}{\m \per \s}$\\
$e$ & The elementary charge, $\SI{1.60e-19}{C}$\\
$h$ & The Planck constant, $\SI{6.63e-34}{J \s}$\\
$\hbar$ & The reduced Planck constant, $\frac{h}{2\pi}$ \\
\end{tabular}
\end{table}

\thispagestyle{plain}

\pagebreak

\pagenumbering{arabic}
\setcounter{page}{1}

\section{Introduction}
\addtocontents{toc}{\protect\thispagestyle{empty}}
\subsection{Motivation}
Graphene is a two dimensional sheet of carbon atoms arranged in a honeycomb lattice. Since its discovery in 2004 by Geim and Novoselov, for which they shared the Nobel prize in 2010 \cite{nobel}, graphene has captured the interests of scientists and engineers alike. Due to its two-dimensional nature, graphene possesses a myriad of novel electronic, mechanical, thermal, and optical properties that make it a potential candidate for several applications including flexible electronics, touch screens, biological and chemical sensing, drug delivery, and transistors \cite{review1, review, biosense, thzdetect, sense}. Indeed, the application of graphene to electronics is now a burgeoning research area, and has come along way since its genesis in 2004. 

The transistor is a key building block of virtually all modern electronic devices. The first transistor was invented in 1947 by Shockley, Bardeen, and Brattain at Bell Labs, and represented a revolutionary advancement in the development of electronic devices in the latter half of the 20th century. Different types of transistors, including bipolar junction transistors (BJTs) and field effect transistors (FETs) were invented in the 20th century -- but the most commonly used transistor in modern electronics is the metal oxide semiconductor field effect transistor (MOSFET), which was invented by Atalla and Kahng in 1959 at Bell Labs. Complementary metal oxide semiconductor (CMOS) technology uses MOSFETs made primarily of silicon, and is the most widely used technology for realizing electronics today \cite{sedra, basics}. 

Since its inception, physicists and engineers have downscaled the size of the MOSFET transistor while maintaining its performance, which has been the driving force behind the incredible speed at which technology has progressed over the past few decades. In more concrete terms, this is described by Moore's law. Moore's law is the observation that the number of transistors on an integrated circuit (and, in turn, computer processing power) doubles every two years at the same cost as a result of downscaling the size of the MOSFET transistor \cite{gordon, moore1, moore2}. Figure 1 depicts the number of transistors on computer chips as a function of time, from 1965 to 2015, which can be seen to roughly vary according to Moore's law. As of 2019, the number of transistors on commercially available microprocessors can reach up to $39.54$ billion \cite{score}, and Samsung and TSMC have been fabricating $\SI{10}{nm}$ and $\SI{7}{nm}$ MOSFETs \cite{samsung, TSMC}. 

\begin{figure}[H]
\begin{center}
\includegraphics[scale=0.4]{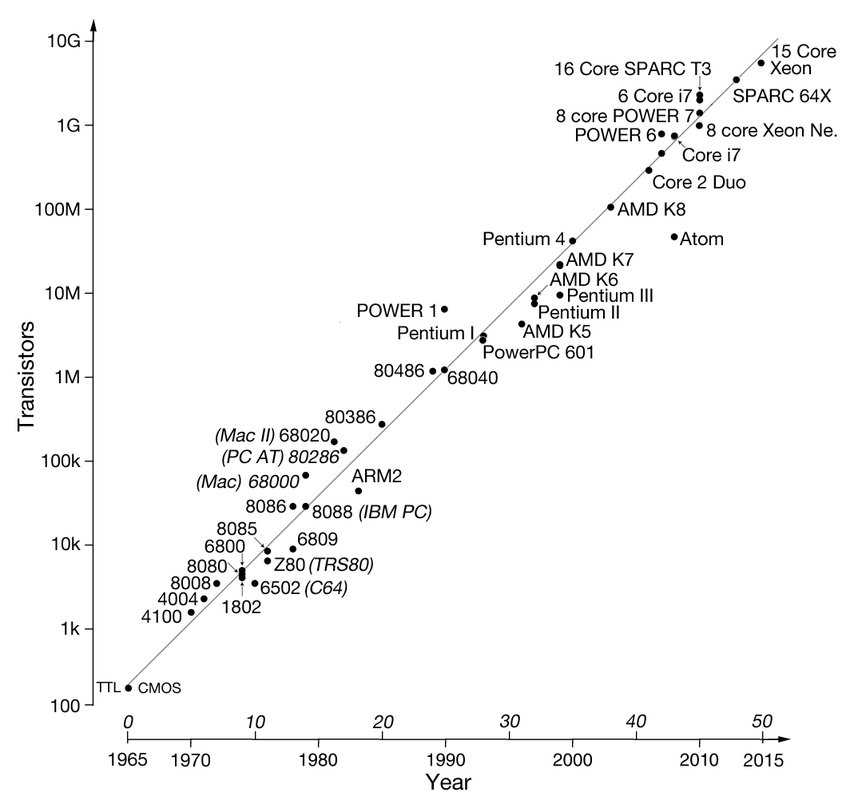}
\caption{A graphical illustration of Moore's law from 1965 to 2015. The vertical axis, showing the transistor count, is logarithmic. It is evident that, to a good approximation, the number of transistors on a computer chip has doubled every two years, for the past five decades \cite{moore1}.}
\label{fig:moore}
\end{center}
\end{figure}

Recently, however, it has been observed that Moore's law is beginning to reach a plateau, as the miniaturization of transistors continues, and is predicted to end around 2025 \cite{moorend}. Moreover, the International Technology Roadmap for Semiconductors predicts that after the year 2021, downscaling transistors will no longer be economically viable \cite{shrink}. This is primarily because at small scales, undesirable short-channel effects such as drain induced barrier lowering, velocity saturization, impact ionization, and other quantum mechanical phenomena begin to manifest, degrading MOSFET performance \cite{short}. As such, physicists and engineers are considering alternative avenues and technologies for extending Moore's law in a post-silicon world. Among the chief novel materials that provide a way of achieving this goal is graphene \cite{review1}. 

The remarkable electronic properties exhibited by graphene, including its extraordinarily high mobility and its ambipolar field effect behavior, make it a promising candidate for carrying electric current in FETs and could in principle outperform existing silicon-based technologies \cite{review, review1}. Since 2007, efforts have been made toward incorporating graphene into existing MOSFET technology \cite{dissertation}. These graphene-based FETs have a number of important potential engineering applications, including sensors \cite{sense, sense2} and high frequency terahertz detectors \cite{thzapp}. The latter is of particular importance in engineering due to the so-called ``terahertz gap" --  a region in the electromagnetic spectrum extending roughly from  $\SI{0.1}{T Hz}$ to $\SI{10} {THz}$ for which existing generation/detection technologies are inadequate. Terahertz technology has a number of potential applications including medical imaging, security scanning, and as a tool in molecular biology research \cite{thzapp, thz1, thz2, thzdetect}. However, there exist economic and physical challenges and bottlenecks associated with realizing graphene FETs that are suitable for the aforementioned applications. This report provides a review of the physics of graphene and its electronic properties as relevant in the context of field effect transistors as well as a state-of-the-art review of different graphene FET implementations.

\subsection{Layout of Report}
The remainder of this report is split up into four main sections. In section 2, a brief historical overview of graphene is presented, followed by a review of the physics of graphene with particular emphasis on its crystallography and electronic band structure. Relevant electronic properties, such as the high mobility of graphene and its ambipolar field effect behavior, are described. Different methods of synthesizing graphene are presented and compared in terms of their scalability, cost, and the quality of graphene production. Finally, the topic of bandgap engineering in graphene is discussed, using bilayer graphene and graphene nanoribbons as examples. 

In section 3, the principle of operation of the conventional MOSFET transistor is discussed, and an overview of basic MOSFET device physics is presented. The MOSFET transistor is modelled as a three terminal device, and relevant current-voltage characteristics are highlighted. Key figures of merit that are commonly found in the literature are extracted from the model, and are used in section 4 to compare different graphene FET implementations. 

In section 4, a state-of-the-art review of graphene FETs is presented, with particular focus on monolayer graphene FETs, bilayer graphene FETs, and graphene nanoribbon FETs. Different implementations in the literature are compared using the figures of merit presented in section 3, and the challenges associated with improving the performance of graphene FETs are identified and discussed. 

Finally, in section 5, the key ideas pertaining to the state-of-the-art graphene FETs presented in section 4 are summarized, and an assessment of the current state of graphene FET research within the wider context of modern industrial applications is presented. 

\pagebreak

\section{Graphene}
Graphene is a single atom-thick planar allotrope of carbon. It is closely related to graphite, which is another allotrope of carbon \cite{notes, review1}. The structure of three dimensional graphite, which may be thought of as a layered stack of graphene sheets held together by van der Waals forces, was determined and studied in 1916 through the use of powder diffraction \cite{diffraction}. The difference in the structure of two dimensional graphene and three dimensional graphite is shown in Fig. 2 

\begin{figure}[H]
\begin{center}
\includegraphics[scale=0.7]{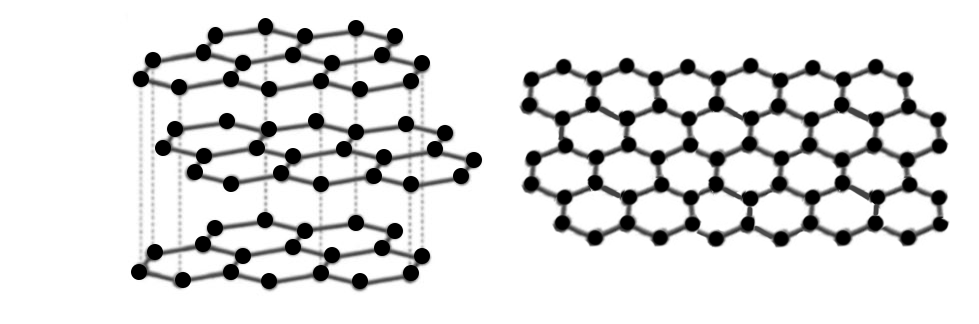}
\caption{A diagram illustrating the difference between graphite (a) and graphene (b). Graphite is made of several layers of graphene sheet stacked on top of one another and held together via weak van der Waals forces \cite{graphitefig}.}
\label{fig:moore}
\end{center}
\end{figure}

The theory of monolayer graphite, or graphene, was not developed until 1947 when Wallace studied the electronic band structure of graphene in order to gain some understanding of the electronic properties of three dimensional graphite by extrapolating the electronic properties of graphene \cite{wallace}. Despite efforts to study the physics of graphene, physicists had long ruled out its existence as a two dimensional crystal in a free state due to the Mermin-–Wagner theorem and the Landau--Peierls arguments concerning thermal fluctations at nonzero temperatures which lead to thermodynamically unstable two dimensional crystals \cite{notes,review}. 

In 2004, at the University of Manchester, Geim and Novoselov demonstrated the first experimental evidence of the existence of graphene by exfoliating crystalline graphite using scotch tape and transferring the graphene layers onto a thin silicon dioxide over silicon \cite{ambipolar, nobel}-- a technique now referred to as mechanical exfoliation. Soon after, the anomalous quantum Hall effect was observed in graphene and reported by Geim and Novoselov as well as Kim and Zhang at Columbia University \cite{anomalous, hall2}. The observation of the anomalous quantum Hall effect provided experimental evidence for the interesting relativistic behavior of electrons in graphene -- in particular, it was shown that electrons in graphene may be viewed as massless charged fermions \cite{anomalous}. As shall be explained in this section, the relativistic behavior of electrons in graphene gives rise to its extraordinary electronic properties.

\subsection{Crystallography and Band Structure}
Graphene has a honeycomb lattice of carbon atoms separated by an interatomic distance $a \approx \SI{1.42}{\angstrom}$ \cite{notes}. Figure 3 shows a scanning tunnelling microscopy (STM) image of graphene that depicts its honeycomb network of carbon atoms. Figure 4 shows a sketch of the honeycomb lattice of graphene and highlights the different environments of neighboring carbon atoms in its lattice.

\begin{figure}[H]
\begin{center}
\includegraphics[scale=0.36]{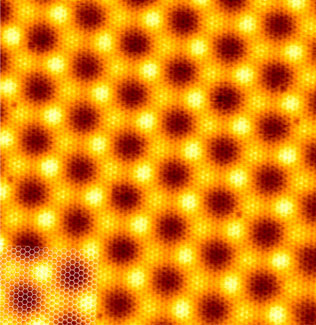}
\caption{An STM image of graphene on the substrate Ir(111) that shows its honeycomb structure \cite{stm}.}
\label{fig:moore}
\end{center}
\end{figure}

\begin{figure}[H]
\begin{center}
\includegraphics[scale=0.55]{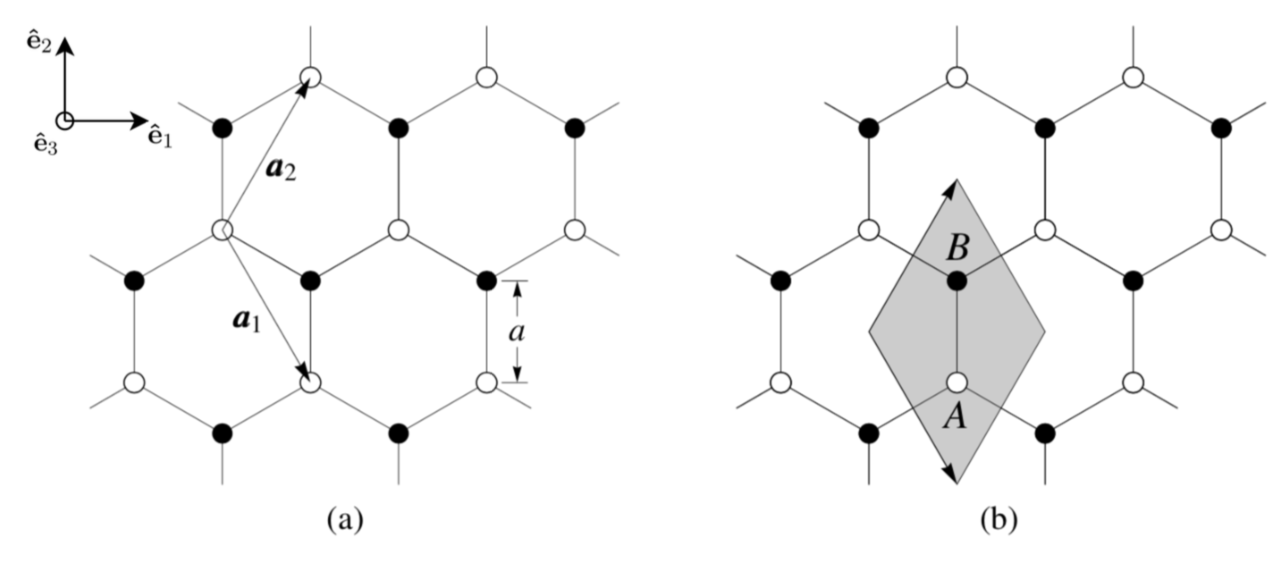}
\caption{(a) The honeycomb lattice, with the primitive lattice vectors $\mathbf{a}_1$ and $\mathbf{a}_2$ that span the lattice in real space. Different carbon atoms corresponding to filled and unfilled circles are not equivalent in crystallographic terms, so the honeycomb lattice is not a Bravais lattice. (b) Shows how this problem can be overcome by defining the shaded region to be a unit cell containing two distinguishable carbon atoms. The two atoms may be thought of as atoms from two different interpenetrating sublattices, labelled $A$ and $B$ \cite{notes}.}
\label{fig:moore}
\end{center}
\end{figure}

As shown in Fig. 4, different atoms in the lattice are not equivalent, making the honeycomb lattice a non-Bravais lattice. These two inequivalent sublattices, labelled $A$ and $B$, may be thought of as interpentrating sublattices that form a triangular Bravais lattice with two atoms per unit cell and two primitive lattice vectors $\mathbf{a}_1$ and $\mathbf{a}_2$ \cite{notes}. With reference to the coordinate system defined by the right-handed orthonormal set of vectors $(\mathbf{\hat{e}}_1, \mathbf{\hat{e}}_2, \mathbf{\hat{e}}_3)$ is such that $\mathbf{\hat{e}}_1$ and $\mathbf{\hat{e}}_2$ lie in the plane of graphene, with  $\mathbf{\hat{e}}_3$ pointing in a direction perpendicular to the plane. The primitive lattice vectors are given by

\begin{equation}
\mathbf{a}_1 = \frac{\sqrt{3} a}{2} (\mathbf{\hat{e}}_1 - \sqrt{3}\mathbf{\hat{e}}_2) = \frac{a'}{2} (\mathbf{\hat{e}}_1 - \sqrt{3}\mathbf{\hat{e}}_2)
\label{eq:prim1}
\end{equation}
and
\begin{equation}
\mathbf{a}_2 = \frac{\sqrt{3}a}{2} (\mathbf{\hat{e}}_1 + \sqrt{3}\mathbf{\hat{e}}_2) = \frac{a'}{2} (\mathbf{\hat{e}}_1 + \sqrt{3}\mathbf{\hat{e}}_2),
\label{eq:prim2}
\end{equation}
where $\left\lVert \mathbf{a}_{1,2} \right\rVert = a' = \sqrt{3} a \approx \SI{2.46}{\angstrom}$ is the lattice constant. The primitive reciprocal lattice vectors, $\mathbf{b}_1$ and $\mathbf{b}_2$, are related to $\mathbf{a}_1$ and $\mathbf{a}_2$ \cite{basics} by 

 \begin{equation}
\mathbf{b}_1 = 2 \pi \frac{\mathbf{a}_2 \times \mathbf{\hat{e}}_3}{\mathbf{a}_1 \cdot (\mathbf{a}_2 \times \mathbf{\hat{e}}_3)} = \frac{2 \pi}{a'} \left( \mathbf{\hat{e}}_1 - \frac{\mathbf{\hat{e}}_2}{\sqrt{3}} \right)
\label{eq:recip1}
\end{equation}
and
\begin{equation}
\mathbf{b}_2 = 2 \pi \frac{\mathbf{\hat{e}}_3\times \mathbf{a}_1}{\mathbf{a}_1 \cdot (\mathbf{a}_2 \times \mathbf{\hat{e}}_3)} = \frac{2 \pi}{a'} \left( \mathbf{\hat{e}}_1 + \frac{\mathbf{\hat{e}}_2}{\sqrt{3}} \right).
\label{eq:recip2}
\end{equation}
\linebreak

\begin{figure}[H]
\begin{center}
\includegraphics[scale=0.55]{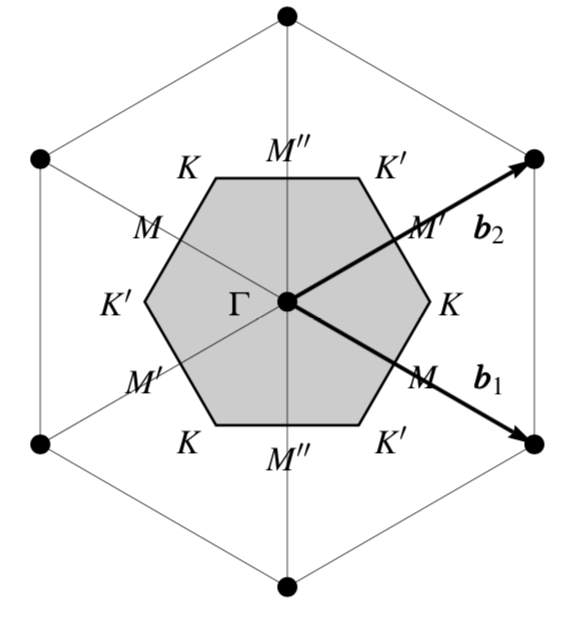}
\caption{The reciprocal lattice of graphene (in $\mathbf{k}$ space). The shaded region is the first Brillouin zone, which is a Wigner--Seitz cell of the reciprocal lattice. By convention, $\Gamma$ denotes the point $\mathbf{k} = \mathbf{0}$. The points $M$ and $M'$, as well as $K$ and $K'$, are inequivalent. Equivalent points are separated in $\mathbf{k}$ space by $\mathbf{b}_1$ and $\mathbf{b}_2$. The corners of the first Brillouin zone, namely, the six points labelled $K$ and $K'$, are collectively referred to as Dirac points \cite{notes}.}
\label{fig:moore}
\end{center}
\end{figure}

Figure 5 shows the first Brillouin zone for graphene in reciprocal space. The center of the first Brillioin zone is labelled as $\Gamma$ by convention and corresponds to the origin $\mathbf{k} =  \mathbf{0}$, where $\mathbf{k} = (k_x, k_y)$ is the wave vector associated with electronic states in the lattice, with $k_x$ and $k_y$ representing the wavenumbers along $\mathbf{\hat{e}}_1$ and $\mathbf{\hat{e}}_2$, respectively. The first Brillouin zone is hexagonal and has six points labelled $K$ and $K'$, collectively referred to as Dirac points. Points with the same label are considered to be equivalent and are separated by a primitive reciprocal lattice vector ($\mathbf{b}_1$ or $\mathbf{b}_2$). The novel electronic properties of graphene hinge on the excitations around these six Dirac points, as shall be explained in this section. It is worth however, showing first the full band structure of graphene computed from first principle calculations. We use VASP to find the full band structure of Graphene. The results are presented in the graph below,
\begin{figure}[h]
    \centering
    \includegraphics[scale=0.6]{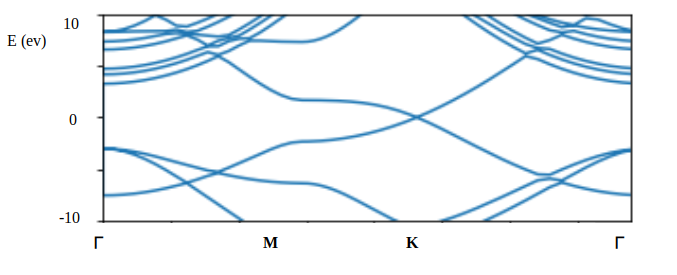}
    \caption{Full Band structure of Graphene calculated using ab initio methods. Here we use VASP to find the Band structure. }
    \label{fig:my_label}
\end{figure}
\newline  The interesting feature of the above band structure as we shall see below is the crossing at point K. This crossing is responsible for many of the interesting properties of graphene. Although the ab initio method susccsfully provides the full band structure with great detail, to understand the physics governing graphene, it is useful to zoom in  at point K and describe the dispersion using a simplified version.
\newline An isolated carbon atom in an excited state has four electrons in its outer shell. Using spectroscopic notation, this corresponds to one $2s$ electron and one electron per $2p$ orbital ($2p_x$, $2p_y$, and $2p_z$). In graphene, the $2s$, $2p_x$, and $2p_y$ states mix to form three $sp^2$ hybrid orbitals for each carbon atom separated by $120^\circ$. The electronic $sp^2$ hybrid states participate in three strong covalent $\sigma$ bonds between each carbon atoms and neighboring carbon atoms in the graphene lattice, leading to the geometry of the lattice shown in Fig. 4. Electrons in $2p_z$ orbitals are located above and below the plane of graphene, participating in weaker $\pi$ bonds \cite{pi}. These electrons will henceforth be referred to as $\pi$ electrons. This is illustrated in Fig. 6.

\begin{figure}[H]
\begin{center}
\includegraphics[scale=0.7]{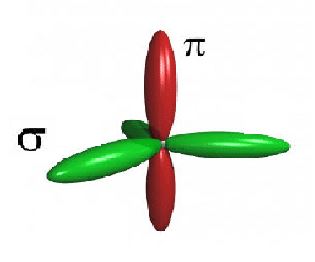}
\caption{A sketch illustrating the different carbon-carbon bonds present in graphene. The electronic states that give rise to the electronic properties of graphene those in the $2p_z$ orbital lobes, labelled $\pi$ on the diagram \cite{pi}.}
\label{fig:moore}
\end{center}
\end{figure}

The $sp^2$ electrons participating in strong $\sigma$ bonds lead to the high strength and other novel mechanical properties of graphene but play no role in the low energy excitations which govern the electronic properties thereof that are relevant in the context of graphene electronics \cite{review1}. On the other hand, the $\pi$ electrons are highly mobile and play a crucial role in the context of the electronic properties of graphene. For this reason, the band structure of graphene as presented and analyzed in the literature only takes into account $\pi$ electrons, which will be assumed in the remainder of this report. 

By applying the tight binding model \cite{notes}, it can be shown (derived in the appendix) that the analytical expression for the energy dispersion relation of $\pi$ electrons is 

\begin{equation}
\epsilon^{(\pm)}(\mathbf{k}) = \frac{\epsilon_0 \pm t\, f(\mathbf{k}) }{1 \pm s\, f(\mathbf{k})},
\label{eq:band}
\end{equation}
where $\epsilon = \epsilon^{(\pm)}(\mathbf{k})$ is the energy, $\epsilon_0$ is a parameter that sets the zero of the dispersion relation, $t$ is a tight binding hopping parameter, $s$ is an overlap parameter, $+$ and $-$ denote the valence and conduction bands respectively, and $f$ is a function defined by 

\begin{equation}
f(\mathbf{k}) = \sqrt{1 + 4 \cos{\left( \frac{3k_y a}{2} \right)} \cos{\left( \frac{\sqrt{3}k_x a}{2} \right)} + 4\cos^2{\left( \frac{\sqrt{3}k_x a}{2} \right)}}.
\label{eq:function}
\end{equation}
\linebreak

\begin{figure}[H]
\begin{center}
\includegraphics[scale=0.7]{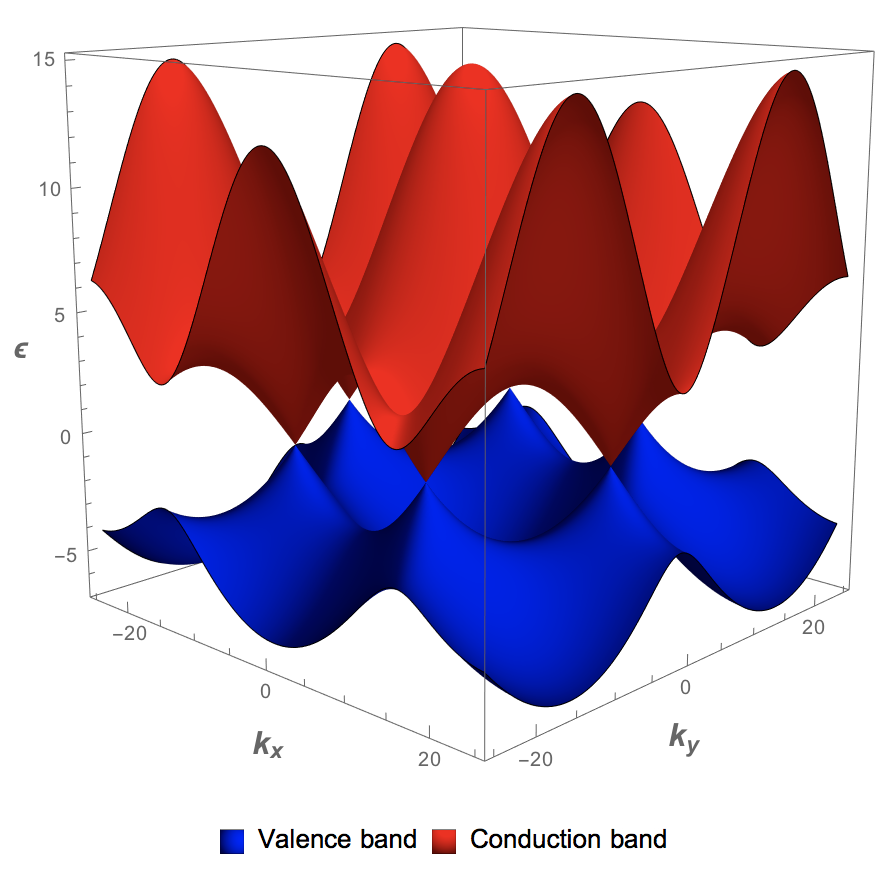}
\caption{A Mathematica plot of the energy dispersion relation of graphene (Eq. (5)), showing the valence (blue) and conduction (red) bands in the first Brillouin zone. The valence and conduction bands touch at six points (the Dirac points) resulting in a zero energy bandgap.}
\label{fig:moore}
\end{center}
\end{figure}

Figure 7 depicts surface plot of the valence (blue) and conduction (red) bands in the first Brillouin zone in accordance with Eq. (\ref{eq:band}), using the values $\epsilon_0 = 0$, $t = \SI{-3.033}{\eV}$, and $s = 0.129$. The values for $t$ and $s$ were obtained from \cite{ts}. The band structure shows that the valence and conduction bands of graphene coincide at six points (the Dirac points of the reciprocal lattice), indicating a zero bandgap\footnotemark. Thus, graphene is semimetallic. The six (Dirac) points at which the valence and conduction bands touch correspond to zeros of the function $f$ (defined in Eq. (\ref{eq:function})) within the first Brillouin zone. The zeros are located at 

\begin{equation}
\mathbf{k} \in\set[\bigg]{\left( \pm \frac{4 \pi}{3 a'}, -\frac{4 \pi}{3 a} \right), \left( \pm   \frac{ 4 \pi}{3 a'}, 0 \right), \left(\pm  \frac{4 \pi}{3 a'}, \frac{4 \pi}{3 a} \right)},
\label{eq:zero}
\end{equation}
where $+$ and $-$ signs distinguish $K$ points from $K'$ points at every value of $k_y$, such that two adjacent points are inequivalent. The zero bandgap of graphene has a number of implications with regards to its use in field effect transistors, as shall be elaborated in later sections.

\footnotetext{It should be noted that the bandgap of graphene is not exactly zero. It has been shown that spin-orbit coupling in graphene can open a small energy bandgap, on the order of $\SI{1}{\micro eV}$ \cite{so}.}

The behavior of the dispersion relation (Eq. (5)) near the Dirac points may be approximated by carrying out first order Taylor expansion of the function $f$ defined by Eq. (6), resulting in the linearized expression 

\begin{equation}
\epsilon^{(\pm)}(\mathbf{k} - \mathbf{k}_0) = \pm \hbar v_F \left\lVert \mathbf{k} - \mathbf{k}_0 \right\rVert ,
\label{eq:cones}
\end{equation}
where $\hbar$ is the reduced Planck constant, $\mathbf{k}_0$ is the wave vector a Dirac point, and $v_F$ is the Fermi velocity of the electrons, given by 

\begin{equation}
v_F = \frac{3a |t|}{2 \hbar} \approx \frac{1}{300} c,
\label{eq:fermi}
\end{equation}
where $c$ is the speed of light in vacuum. Equation (\ref{eq:cones}) indicates that the conduction and valence bands take a conical shape near each Dirac point, forming so-called Dirac cones. In fact, in the vicinity of the Dirac points, this is in agreement with dispersion relation plot Fig. 7, as illustrated in Fig. 8.

\begin{figure}[H]
\begin{center}
\includegraphics[scale=0.7]{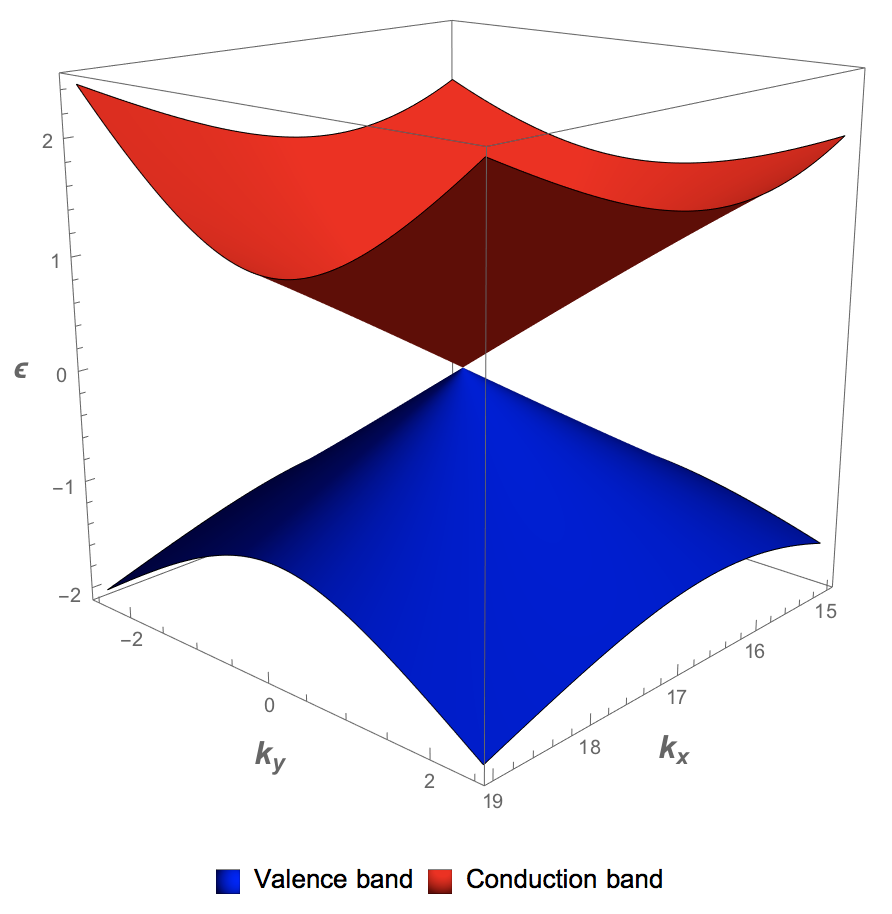}
\caption{A zoomed in version Mathematica plot of the energy dispersion relation of graphene (Eq. (5)), shown in Fig. 7, in the vicinity of one of the Dirac points. It is evident that the dispersion relation becomes approximately conical in the vicinity of the Dirac points, forming Dirac cones.}
\label{fig:moore}
\end{center}
\end{figure}

Another crucial result that appears in the vicinity of the Dirac points is the relativistic Dirac equation. In fact, by finding the first order Taylor expansion of the function $f$, the matrix representation $\mathcal{H}$ of the Hamiltonian describing electrons near the Dirac points may be written as \cite{review}\footnotemark  
\begin{equation}
\mathcal{H} = \hbar v_F \boldsymbol{\sigma} \cdot \mathbf{k}' ,
\label{eq:dirac}
\end{equation}
where $\boldsymbol{\sigma} = (\sigma_x, \sigma_y)$ is a vector of $2 \times 2$ Pauli matrices $\sigma_x$ and $\sigma_y$ given by 

\footnotetext{This equation, with the vector $\mathbf{k}'  \equiv \mathbf{k} - \mathbf{k}_0$, is only valid for the $K$ points of the first Brillouin zone. The equivalent Dirac equation for the $K'$ points may be written in the same form if $\mathbf{k}'$ is redefined such that $k'_x \rightarrow -k'_x$ \cite{notes}.}

\begin{equation}
\sigma_x = \begin{pmatrix}
0  \,\,&\,\,\,\,  1 \\
1  \,\,&\,\,\,\,  0
\end{pmatrix}
\label{eq:pauli1}
\end{equation}
and 

\begin{equation}
\sigma_y = \begin{pmatrix}
0  \,\,&\,  -i \\
i  \,\,&\,  0
\end{pmatrix},
\label{eq:pauli2}
\end{equation}
$\mathbf{k}'  \equiv \mathbf{k} - \mathbf{k}_0$, and $\cdot$ denotes a standard dot product (component-wise multiplication). Before we discuss the physical implication of the above, it is convenient to slightly digress to describe the Dirac physics concluded above. Originally, the Dirac equation was proposed as a relativistic description of spin $1/2$ particles. In a general form, the equation can be given by.
\begin{equation}
    i\hbar \gamma^{\mu} \partial_{\mu}\psi -mc\psi= 0
\end{equation}
where $\psi$ is the spinor of the fermion and $\gamma$ are the so called Gamma matrices. Comparing with equation (13),equation (\ref{eq:dirac}) is the Dirac equation for massless relativistic fermions in two dimensions. Thus, $\pi$ electrons in graphene behave like massless relativistic particles near the Dirac point, making graphene a miniaturized laboratory for testing models from quantum field theory \cite{lab}. Evidently, graphene is a material of great interest not only in the realm of condensed matter physics and electronic engineering research, but also in high energy physics. 
\subsection{Physical Properties}
Graphene exhibits a myriad of remarkable mechanical, optical, and thermal properties in addition to its novel electronic properties \cite{review1}. Some of these properties include high transparency (graphene only absorbs about $2.3 \%$  of visible light), high thermal conductivity (up to \linebreak$\SI{5000}{\W \per \m \per \K}$), and extraordinary mechanical properties (it is simultaneously the strongest and thinnest material ever discovered, with a tensile strength of $\SI{130}{G\Pa}$, about $200$ times stronger than steel) \cite{review, review1}. More information about these properties can be found in \cite{review1}. Instead, the main focus of this section is on the electronic properties of graphene.

It has been reported that graphene possesses a very high intrinsic electron mobility, ideally exceeding $\SI{2e5}{\cm\squared \per \V \per \s}$ at room temperature \cite{review1, dissertation}. In fact, recently, it has been reported that heterostructures made of $\mathrm{WSe_2}$, graphene and $\mathrm{hBN}$ exhibit mobilities as high as $\SI{3.5e5}{\cm\squared \per \V \per \s}$ \cite{350}. Graphene is capable of carrying large currents, with an electrical conductivity higher than that of silver and zinc \cite{review1}. The high mobility of graphene is in large part due to Eq. (\ref{eq:dirac}) which implies that electron backscattering is suppressed \cite{review}. Another explanation for the high mobility of graphene is that it exhibits weak acoustic electron-phonon interactions \cite{review1}.

Graphene has a density of states $g = g(\epsilon)$ given by 

\begin{equation}
g(\epsilon) = \frac{8\pi |\epsilon|}{h^2 v_F},
\label{eq:dos}
\end{equation}
where $\epsilon$ is the energy, $h$ is the Planck constant, and $v_F$ is the Fermi velocity defined in Eq. (\ref{eq:fermi}) \cite{hall2}. Therefore, the density of states of graphene is zero at the Dirac points. However, graphene possesses a minimum conductivity, $\sigma_0$, on the order of $4e^2/h$, where $e$ is the elementary charge and $h$ is the Planck constant \cite{hall2}. This is in accordance with the experimentally observed anomalous quantum Hall effect in graphene -- the Hall conductivity, $\sigma_{xy}$, of graphene was found to be related to the Landau level $N\in\mathbb{Z}$ and the minimum conductivity $\sigma_0$ \cite{hall2,anomalous} by

\begin{equation}
\sigma_{xy} = \frac{4 N e^2}{h} + \frac{1}{2} \sigma_0.
\label{eq:hall}
\end{equation}

Although the mobility of suspended graphene can exceed $\SI{2e5}{\cm\squared \per \V \per \s}$ in ideal cases, there seems to be some variability in observed mobilities in graphene samples; for example, mobilities ranging from $\SI{1e3}{\cm\squared \per \V \per \s}$ to $\SI{7e4}{\cm\squared \per \V \per \s}$ for graphene over $\mathrm{SiO_2}$ \cite{review, dissertation}. This is because mobility can be limited by temperature, defects, and substrate-induced corrugations \cite{review, review1}. Thus, different synthesis techniques result in graphene samples with different mobilities. Figure 9 shows how mobilities observed in different graphene samples vary as a function of charge carrier density fluctuation induced by disorder in each sample \cite{350}.

\begin{figure}[H]
\begin{center}
\includegraphics[scale=0.5]{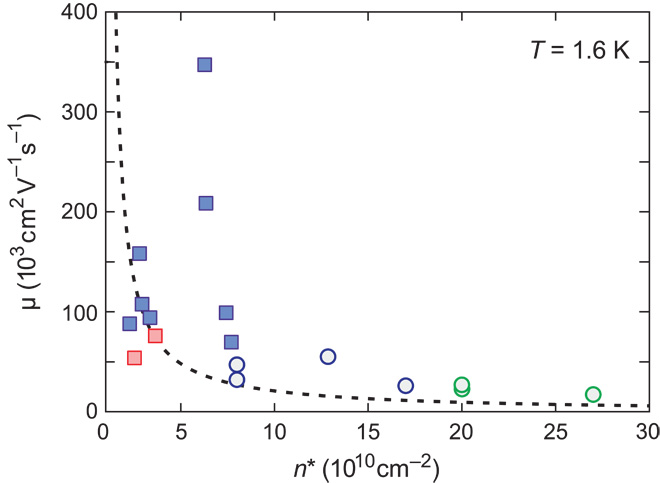}
\caption{A plot of mobility $\mu$ as a function of disorder-induced charge carrier density fluctuation $n^*$ (at a temperature of $\SI{1.6}{K}$) for graphene samples from different studies, compiled by \cite{350}. Samples with $\mathrm{hBN}$ (blue circles and squares), copper (red squares), and $\mathrm{SiO_2}$ (green circles) substrates are considered. It can be seen that large disorder-induced charge carrier densities degrade the mobility of graphene \cite{350}.}
\label{fig:moore}
\end{center}
\end{figure}

Another noteworthy phenomenon that was observed in graphene by Geim \etal \, is the so-called ambipolar electric field effect \cite{ambipolar}. It was found that when an electric field, corresponding to a gate voltage, $V_g$, is applied to a sample of exfoliated graphene on a silicon dioxide over silicon substrate, the conductivity $\sigma = \sigma(V_g)$ exhibits a characteristic ``V" shape dependence as shown in Fig. 10. The conductivity is varies linearly in the vicinity of the point of minimum conductivity (which corresponds to the point of charge neutrality), which is on the order of $4e^2/h$, at $V_g = V_{g,\mathrm{min}}$, the point of minimum conductivity. To the right of the minimum, when $\partial \sigma/ \partial V_g > 0$, the majority carries are electrons in the conduction band (graphene is $n$--type), while to the left of the minimum, when $\partial \sigma/ \partial V_g < 0$, the majority carriers are holes in the valence band (graphene is $p$--type). Therefore, graphene can conduct electrons or holes, with a tunable conductivity that varies as a function of the applied gate voltage \cite{review1, ambipolar}. Furthermore, unlike silicon and other semiconductors, electron and hole mobilities in ideal graphene that is free from impurities are nearly equal, as a consequence of the symmetry of Eq. (\ref{eq:cones}) for the valence and conduction bands \cite{equal}.

\begin{figure}[H]
\begin{center}
\includegraphics[scale=0.7]{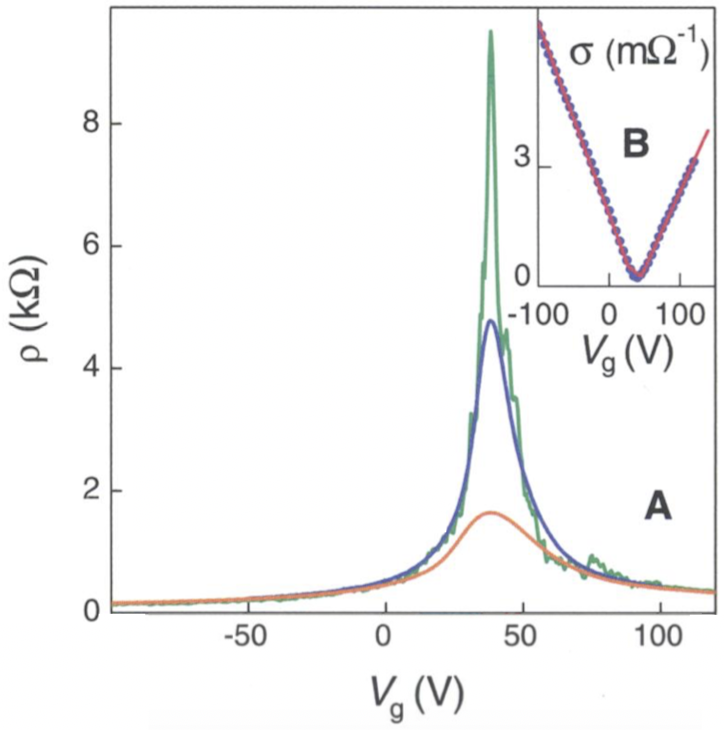}
\caption{Experimental data produced by Geim \etal \, that depicts the ambipolar field effect in graphene. The different colors indicate different temperatures. The conductivity, $\sigma$, is the reciprocal of the resistivity, $\rho$. Therefore, the peak in the resistivity plot corresponds to a minimum in conductivity plot, at $V_g = V_{g,\mathrm{min}}$ \cite{ambipolar}. }
\label{fig:moore}
\end{center}
\end{figure}

It is worth mentioning that conventional doping, whereby atoms in the lattice of a semiconductor (such as silicon or germanium) are replaced by dopant atoms is not possible in the case of graphene due to the strong carbon--carbon covalent bonds in the lattice. Instead, doping in the context of graphene in the literature refers to placing dopant atoms on the surface of graphene, without replacing carbon atoms. The introduction of a dopant atom alters the electronic band structure of graphene and can create a nonzero bandgap. Dopants that are commonly used for graphene in research include boron nitride, sulfur, and gold \cite{review}.

\subsection{Synthesis Techniques}
The most well known (and the oldest) technique for synthesizing graphene is mechanical exfoliation (also referred to as mechanical cleavage in the literature, or, less formally, the ``scotch tape" method). In fact, this was the technique used by Geim \etal \, in 2004 when they isolated graphene layers on thin $\mathrm{SiO_2 / Si}$. The main steps of the process are as follows. A small piece of graphite is obtained from a larger graphite sample. Typically, the graphite sample used in the process is highly ordered pyrolytic graphite (HOPG). The small piece of graphite is then stuck to the surface of an adhesive tape, which is used to peel graphene flakes from the graphite sample by repeatedly folding and unfolding the tape. The graphene layers are then transferred onto the surface of a smooth substrate, such as $\mathrm{SiO_2 / Si}$. and can be verified and located by observing light interference patterns using an optical microscope. \cite{review, ambipolar, review1}. This process is illustrated in Fig. 11.

\begin{figure}[H]
\begin{center}
\includegraphics[scale=1]{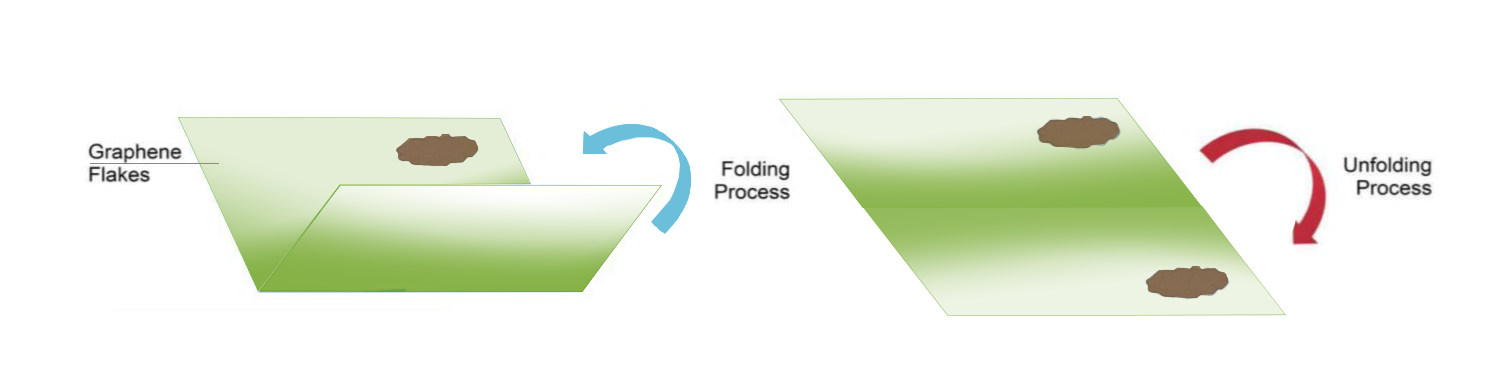}
\caption{A sketch of the mechanical exfoliation process. The adhesive tape is folded and unfolded, gradually peeling off graphene flakes \cite{review}.}
\label{fig:moore}
\end{center}
\end{figure}

The advantage of mechanical exfoliation is that it produces high quality graphene with high mobility and low defects, with the highest recorded mobility exceeding $\SI{2e5}{\cm\squared \per \V \per \s}$ at room temperature. The main drawback of this method is that it is not scalable, and it produces relatively small quantities of graphene -- thus, it is not suitable for industrial applications \cite{review, dissertation}. 

Another method for synthesizing graphene is vacuum epitaxial growth over $\mathrm{SiC}$. In this process, a silicon wafer is coated with $\mathrm{SiC}$. and heated to high temperatures, up to and exceeding $\SI{1100}{\celsius}$, in ultra-high vacuum. At these temperatures, the silicon atoms begin to evaporate while carbon atoms remain, leaving epitaxially grown graphene layers on the substrate \cite{review, dissertation}. This is shown in Fig. 12.

\begin{figure}[H]
\begin{center}
\includegraphics[scale=0.7]{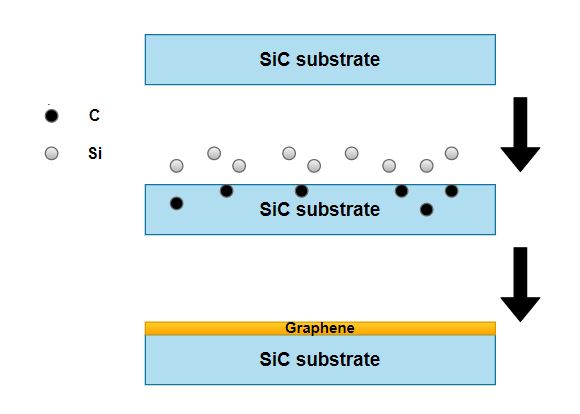}
\caption{An illustration showing the main steps of epitaxial growth over $\mathrm{SiC}$. The high temperature, exceeding, $\SI{1100}{\celsius}$ causes silicon to sublime \cite{epitaxydiagram}.}
\label{fig:moore}
\end{center}
\end{figure}

This technique can produce graphene samples with a mobility of up to $\SI{5e3}{\cm\squared \per \V \per \s}$ at room temperature. It has also been shown that a mobility exceeding $\SI{1.1e4}{\cm\squared \per \V \per \s}$ can be achieved after eliminating dangling silicon bonds from the sample. Epitaxy inevitably results in lower mobility and higher structural defects than mechanically cleavage due to the burning of carbon at high temperatures, which leads to the sample being contaminated by hydrogen and oxygen atoms. However, the technique offers more scalability than mechanical exfoliation \cite{review, dissertation}.

The most commonly used technique in industry for synthesizing graphene is chemical vapor deposition (CVD). This technique involves mixing hydrogen and a gaseous source of carbon such as $\mathrm{CH_4}$ or $\mathrm{C_2H_2}$ over a catalytic bed made of copper or nickel in a chamber. At high temperatures (in excess of $\SI{1000}{\celsius}$), the catalyst breaks the bonds in the gaseous sources and the hydrogen is burned, leaving graphene deposits on the surface of the catalytic bed. This process is illustrated in Fig. 13 \cite{review, dissertation}.

\begin{figure}[H]
\begin{center}
\includegraphics[scale=1.3]{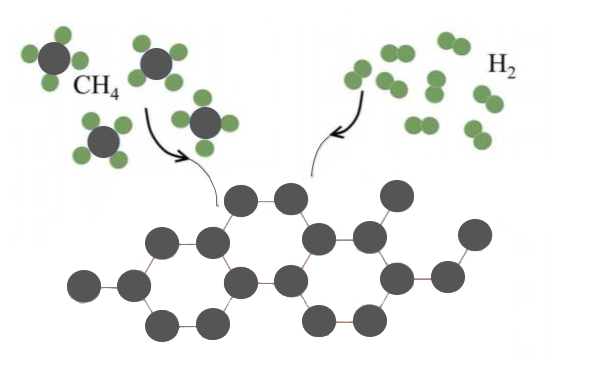}
\caption{An illustration of how graphene is grown using CVD. The carbon--hydrogen bonds in $\mathrm{CH_4}$ are broken at high temperatures over the catalytic bed, and the hydrogen burns and evaporates, leaving graphene deposits on the surface of the bed \cite{review}.}
\label{fig:moore}
\end{center}
\end{figure}

A larger graphene yield can be produced by using a larger catalytic bed. This makes CVD more scalable than other graphene synthesis techniques. In addition, the cost of CVD is lower than that of vacuum epitaxial growth and mechanical exfoliation. This makes CVD more suitable than other techniques in industry. The disadvantage of using CVD for graphene synthesis is the presence of point defects, grain boundaries, and surface contaminants in the yield, all of which typically result in lower mobilities than graphene sample produced via epitaxy or exfoliation \cite{review, dissertation}. However, recently, it was reported that with appropriate cleaning and encapsulation, the room temperature mobility of CVD grown graphene can exceed $\SI{7e4}{\cm\squared \per \V \per \s}$, which is higher than room temperature mobilities observed in epitaxially grown graphene samples \cite{cvdmob}. 

\subsection{Related Structures and Bandgap Engineering}
The model of graphene presented thus far is a two dimensional single layer of carbon atoms in a honeycomb lattice of infinite spatial extent. Before discussing graphene FETs, it is important to explore other structures that are related to the model of graphene discussed in sections 2.1--2.3. The zero bandgap of graphene is undesirable in the context of digital electronics, as shall be elaborated in section 3. Thus, ``opening up" the bandgap of graphene and tuning it is highly desirable for developing graphene FETs. It was previously stated that adding dopants to graphene can result in a nonzero bandgap. However, bandgaps generated via doping are generally not easily tunable \cite{review}. Evidently, bandgap engineering in graphene is crucial, and is an active ongoing area of research. The structures presented in this section offer alternative means of generating bandgaps in graphene. There is, however, a tradeoff -- as these structures exhibit lower mobilities than monolayer graphene. 
\subsubsection{Bilayer Graphene}
As its name suggests, bilayer graphene is a structure that is made of two stacked graphene monolayers held together by van der Waals forces. Figure 14 depicts two forms of bilayer graphene: AB Bernal stacked form, and the less common and more unstable AA stacked form. The two forms differ in the position of one of the graphene sheets relative to the other \cite{aaab}. All the examples of bilayer graphene presented in this report are in AB Bernal stacked form. 

\begin{figure}[H]
\begin{center}
\includegraphics[scale=1]{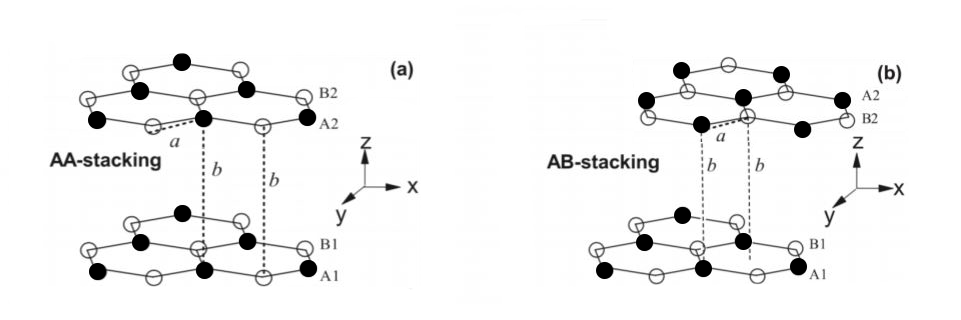}
\caption{(a) AA stacked bilayer graphene, in which the two graphene sheets are exactly aligned. (b) AB stacked bilayer graphene, in which the two graphene sheets are stacked such that half of the atoms in the upper sheet lie over the center of a hexagon in the lower sheet, while the other half are aligned over atoms in the lower sheet \cite{aaab}.}
\label{fig:moore}
\end{center}
\end{figure}

By using an effective two-band Hamiltonian model the following energy dispersion relation for $\pi$ electrons in AB Bernal stacked bilayer graphene can be derived \cite{formula}

\begin{equation}
\epsilon^{(\pm)}(\mathbf{k}) = \frac{U_1 + U_2}{2} \pm \sqrt{\frac{\gamma^2}{2} +  \frac{U^2}{4} + v_F^2 k^2 \hbar^2 - \sqrt{\frac{\gamma^4}{4} + v_F^2 k^2 \hbar^2 (\gamma^2 + U^2)} },
\label{eq:band2}
\end{equation}
where $\epsilon = \epsilon^{(\pm)}(\mathbf{k})$ is the energy, $k = \left\lVert \mathbf{k}\right\rVert$, $U_1$ and $U_2$ are the electrostatic potential energies of the two layers, $U = U_1 - U_2$ is the potential energy difference, $\gamma$ is the interlayer coupling, $v_F$ is the Fermi velocity, and $+$ and $-$ denote the valence and conduction bands respectively. 

Bilayer graphene has an electronic band structure that is different from monolayer graphene; in the vicinity of the Dirac points, the dispersion relation takes a parabolic form, as opposed to the linear/conical form exhibited by monolayer graphene as described by Eq. (\ref{eq:cones}) \cite{dissertation, review1}. In particular, this implies that carriers in bilayer graphene are massive in the vicinity of the Dirac points, as opposed to monolayer graphene, where they behave like massless charged fermions governed by Eq (\ref{eq:dirac}). Bilayer graphene, like monolayer graphene, possesses a zero energy bandgap, when the potential energy difference $U$ between the two layers is zero. However, unlike monolayer graphene, a bandgap can be generated in bilayer graphene by applying an electric field perpendicular to the structure. Furthermore, it was found that the magnitude of the bandgap can be controlled by varying the magnitude of the applied electric field. In particular, it can be shown \cite{formula} that, in accordance with the model used to derive Eq. (\ref{eq:band2}), AB Bernal stacked bilayer graphene has a bandgap, $\Delta$, given by 

\begin{equation}
\Delta = \frac{\gamma |U|}{\sqrt{\gamma^2 + U^2}},
\label{eq:bandgap2}
\end{equation}
which is nonzero for nonzero $U$; \ie, applying a perpendicular electric field generates a nonzero potential energy difference, $U$, between the two layers, opening a bandgap, $\Delta$. It was theoretically shown that, at room temperature, the bandgap of bilayer graphene varies can vary up to $\SI{300}{m\eV}$, and bandgaps up to $\SI{130}{m\eV}$ have been demonstrated \cite{dissertation}. 

Figure 15 shows the approximately parabolic energy dispersion of bilayer graphene near the Dirac points as well as the characteristic ``Mexican hat" shape of the bands when a bandgap $\Delta$ given by Eq. (\ref{eq:bandgap2}) is opened.  

\begin{figure}[H]
\begin{center}
\includegraphics[scale=0.7]{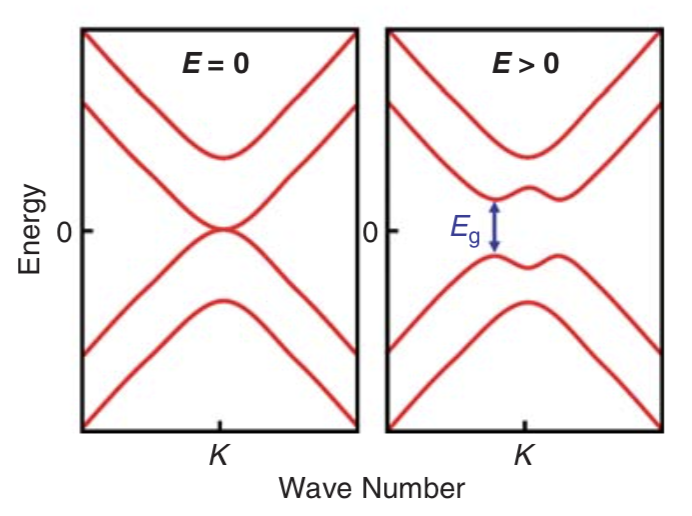}
\caption{A plot of the energy dispersion relation in the vicinity of a Dirac point for AB stacked bilayer graphene in the absence (left) and the presence (right) of an applied perpendicular electric field of magnitude $E$. In this diagram, the bandgap is denoted by $E_g$; whereas in the main text it is denoted by $\Delta$. The dispersion relation is approximately parabolic in the absence of an applied electric field, and shows a characteristic  characteristic ``Mexican hat" shape when a bandgap is opened via the application of a perpendicular electric field \cite{review1}.}
\label{fig:moore}
\end{center}
\end{figure}

Another way of generating a bandgap in bilayer graphene is via doping -- although, as previously stated, bandgaps generated by doping are less tunable \cite{review}. In addition to providing a means of bandgap engineering, bilayer grahene shows low current leakage, which is desirable for graphene FET applications \cite{blg3}. However, these advantages come at the expense of lower carrier mobilities than in monolayer graphene, as theoretically predicted by Wallace \cite{wallace}.

\subsubsection{Graphene Nanoribbons}
A graphene nanoribbon (GNR) is a terminated monolayer graphene sheet of small transverse width on the order of $\SI{50}{nm}$ or less, much smaller than its longitudinal length \cite{review1, gnr}. $\pi$ electrons in GNRs are also governed by Eq. (\ref{eq:dirac}), with different boundary conditions that depend on the edges and geometry of the GNR structure. In particular, the boundary conditions of the Dirac equation can lead to either conducting or semiconducting behavior \cite{review}. There are two variants of GNRs - those with so-called ``armchair" edges and those with ``zigzag" edges, as illustrated in Fig. 16. In particular, zigzag GNRs behave like conductors with no bandgap, whereas armchair GNRs possess a bandgap, $\Delta$, due to confinement of electrons within the structure, and can behave like conductors or semiconductors depending on the number of carbon atoms within the width of the GNR \cite{review}. 

\begin{figure}[H]
\begin{center}
\includegraphics[scale=0.8]{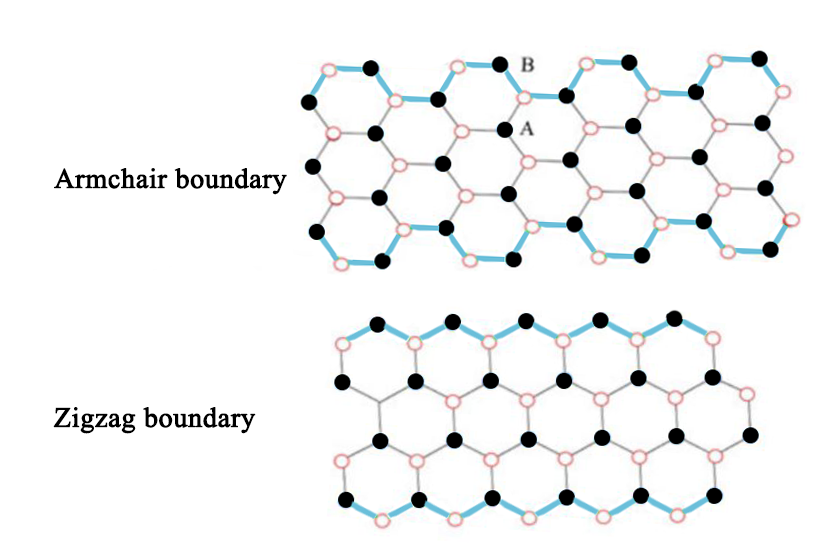}
\caption{An armchair GNR (top) and a zigzag GNR (bottom) \cite{review}.}
\label{fig:moore}
\end{center}
\end{figure}

The energy dispersion relation of an armchair GNR is approximately parabolic \cite{review1} in the vicinity of the Dirac points, with a bandgap $\Delta$ separating the valence and conduction bands, as shown in Fig. 17. 

\begin{figure}[H]
\begin{center}
\includegraphics[scale=0.7]{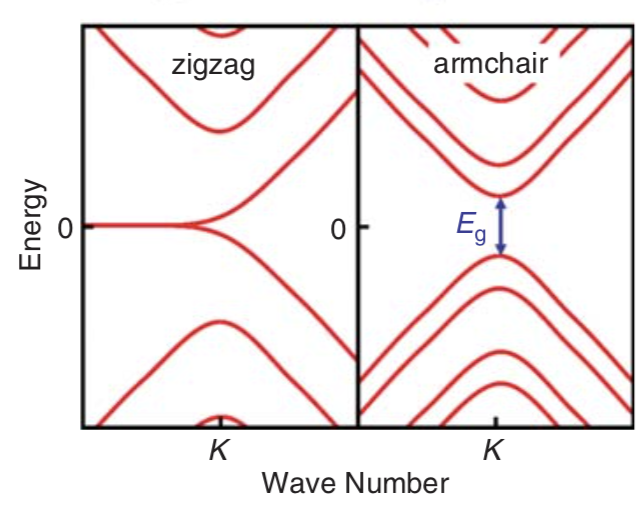}
\caption{A plot of the energy dispersion relation in the vicinity of a Dirac point for zigzag GNR (left) and an armchair GNR (right). In this diagram, the bandgap is denoted by $E_g$; whereas in the main text it is denoted by $\Delta$. The zigzag GNR has a bandgap of zero and shows conducting behavior. On the other hand, the armchair GNR has a bandgap that depends on its width \cite{review1}.}
\label{fig:moore}
\end{center}
\end{figure}

It has been theoretically shown that the bandgap of a GNR with armchair edges is inversely proportional to its transverse width \cite{review1, gnr, review};

\begin{equation}
\Delta \propto \frac{1}{W},
\end{equation}
where $W$ is the width of the GNR. In fact, bandgaps up to $\SI{2.3}{eV}$ have been demonstrated in GNRs \cite{rrr}. However, this dependence does not generally hold in experiments as GNR samples usually contain a mixture of armchair and zigzag edges. Moreover, GNR structures exhibit lower mobilities than monolayer graphene, due to phonon scattering near the edges \cite{review1, review}. Generally, synthesizing well-defined GNRs is a challenging task \cite{gnrmulti}. 

The structure of a GNR is closely related to that of the carbon nanotube (CNT), which is another allotrope of carbon. As depicted in Fig. 18, a CNT has a cylindrical structure of small radius, which may be topologically thought of as a a rolled up GNR \cite{cnt}. In fact, one of the ways in which GNRs can be fabricated is by chemically unzipping carbon nanotubes \cite{review}. Other ways of fabricating GNRs include electron beam lithography and chemical exfoliation of graphite \cite{review1}. 

\begin{figure}[H]
\begin{center}
\includegraphics[scale=0.7]{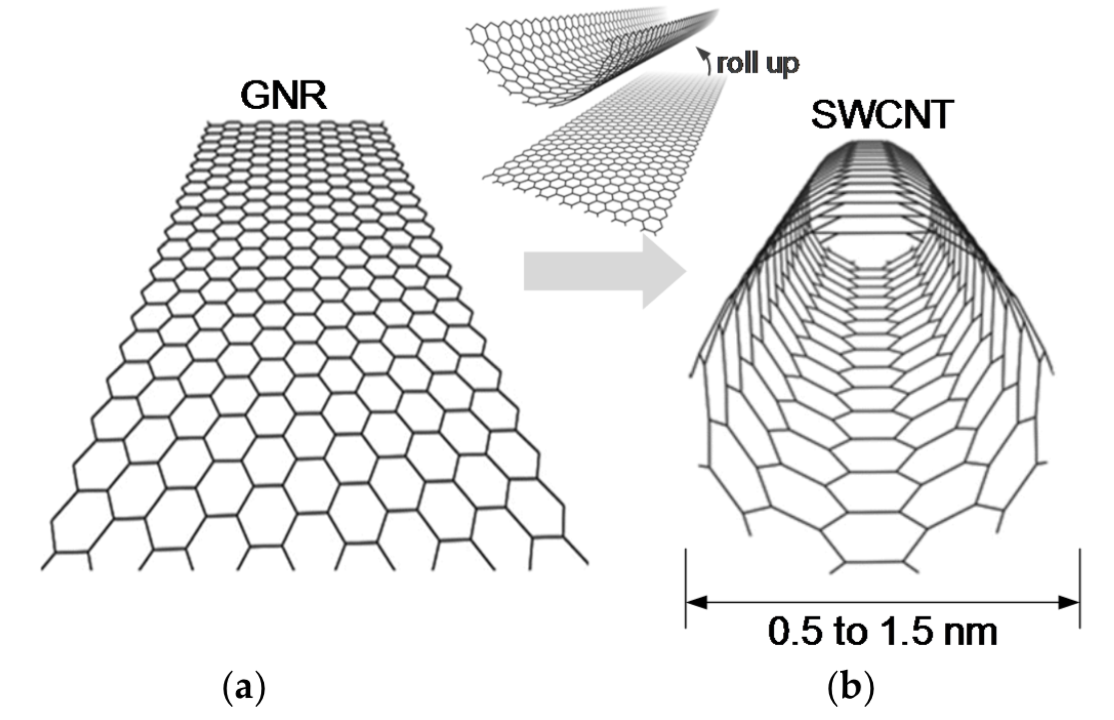}
\caption{(a) An illustration of a GNR. (b) An illustration of a single-walled carbon nanotube (SWCNT). Evidently, the CNT can be thought of as a rolled up GNR \cite{cnt}.}
\label{fig:moore}
\end{center}
\end{figure}

\pagebreak

\section{Conventional CMOS Technology}
MOSFETs have a number of advantages when compared to BJTs, including smaller size and lower power consumption \cite{sedra}. As shall be explained in this section, the MOSFET serves two functions: it can be used as a switch or as an amplifier. The former is used to realize logic gates and digital electronics, while the latter is used to realize analog electronics. MOSFETs of different types can be combined on a single chip to form what is called complementary metal oxide semiconductor (CMOS) technology, which is the chief way in which logic gates and logic operations are implemented in modern integrated circuits (ICs) \cite{moore2}. 

A MOSFET is a semiconducting device with three terminals called the gate, source, and drain \cite{sedra}. This section only describes $n$--channel MOSFETs, but the principles of operation of a $p$--channel MOSFETs are the same. The cross section of an $n$--channel MOSFET and its associated circuit schematic are shown in Fig. 19 and Fig. 20, respectively.\footnotemark

\footnotetext{Note that in diagrams adopted from electrical engineering textbooks (such as \cite{sedra}), the convention of using lower case letters to denote circuit variables is used. This is avoided in the main text, so as not to confuse the current variable $i$ with the imaginary unit $i = \sqrt{-1}$ Thus, voltages and currents are denoted by upper case letters in this report.}

\begin{figure}[H]
\begin{center}
\includegraphics[scale=0.6]{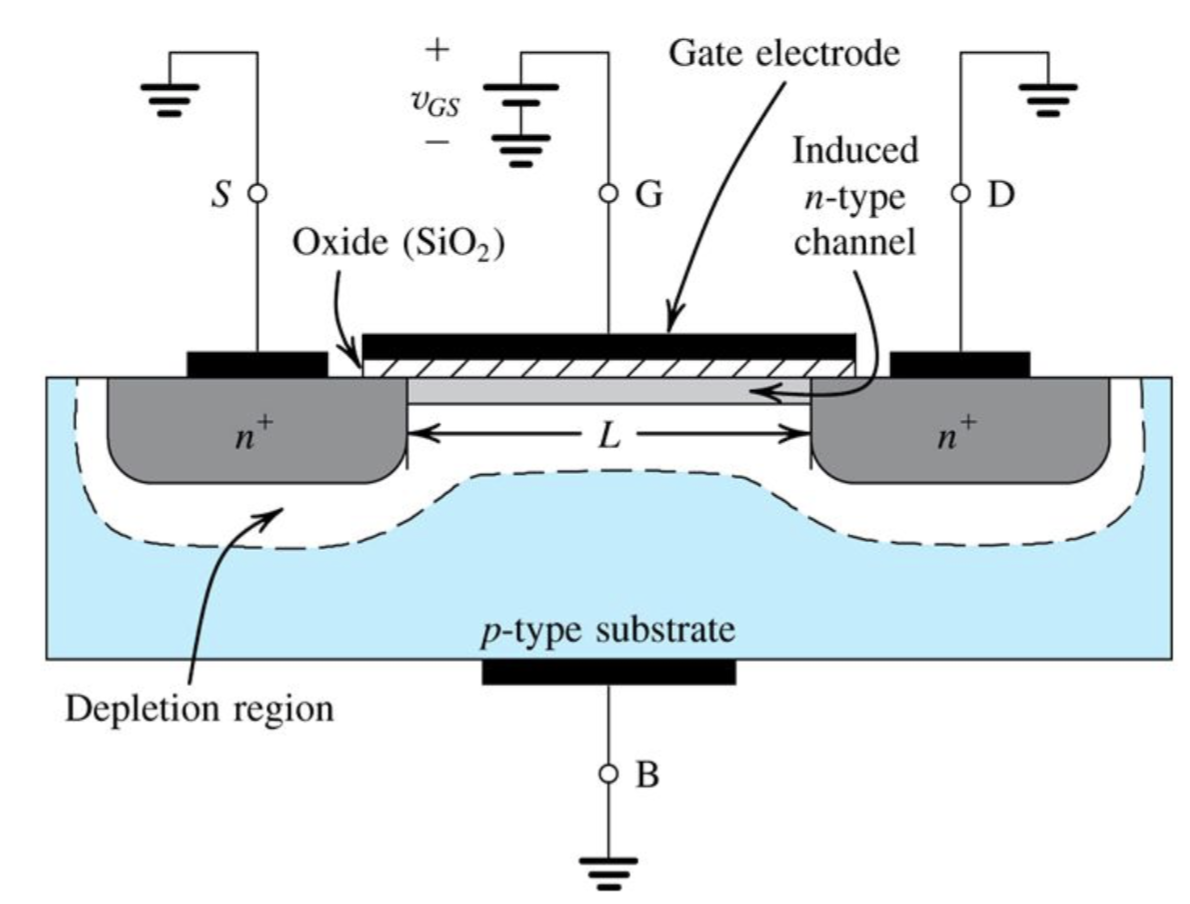}
\caption{Cross-sectional schematic of an $n$--channel MOSFET. In this particular setup, the source (S) and drain (D) terminals are grounded. This is not a general requirement \cite{sedra}.}
\label{fig:moore}
\end{center}
\end{figure}

\begin{figure}[H]
\begin{center}
\includegraphics[scale=0.5]{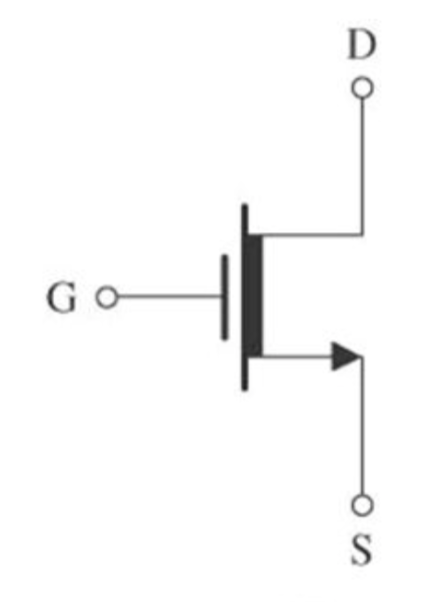}
\caption{Circuit symbol for an $n$--channel MOSFET, showing the gate (G), drain (D), and source (S) terminals \cite{sedra}.}
\label{fig:moore}
\end{center}
\end{figure}

The body of the MOSFET, called the substrate, is a $p$--doped silicon wafer. Two regions on the subtrate are heavily $n$--doped. These regions are referred to as the drain and source regions. On top of the substrate is thin layer of silicon dioxide dielectric of thickness $t_{ox}$ on the order of a few nanometers, covering the region between the source and the drain. Metal contacts are deposited on the source and drain regions, and a layer of metal or polysilicon is added on top of the oxide layer, forming what is called the gate electrode. This defines the source, drain, and gate terminals of the device. The region between the source and the drain is called the channel region, and has length, $L$, and width, $W$. The channel length n most MOSFETs is on the order of tens of nanometers, while the width is typically in the range of $\SI{0.2}{\micro m}$ to $\SI{100}{\micro m}$ \cite{sedra, moore2}.

Suppose that the substrate, source, and gate terminals are grounded. Then, ideally, to back-to-back p--n junctions are formed between the drain and the source, and no current flows when a voltage $V_{DS}$ is applied to the drain. This is called the cutoff region of the MOSFET. When a voltage $V_{GS} > 0$ is applied to the gate, the holes in the $p$--doped substrate are repelled, forming a depletion region beneath the gate, source, and drain terminals, as shown in Fig. 19. Furthermore, majority carrier electrons from the heavily doped $n$--type drain and source regions are attracted to the region underneath the gate, forming an $n$--channel, or an inversion layer. The voltage at which sufficient mobile electrons form in the $n$--channel is referred to as the threshold voltage, $V_{TH}$. When $V_{GS} > V_{TH}$, the MOSFET is switched on, and applying a voltage $V_{DS} > 0$ causes a current to flow from the source to the drain \cite{sedra, moore2}.

When the voltage $V_{DS}$ is less than the so-called overdrive voltage $V_{OV} \equiv V_{GS} > V_{TH}$, the MOSFET is said to be in the triode region, and the drain-source current $I_{DS}$ takes the form 

\begin{equation}
I_{DS} = \mu_n C_{ox} \frac{W}{L} \left( V_{OV} V_{DS} - \frac{1}{2} V_{DS}^2 \right),
\label{eq:triode}
\end{equation}
where $\mu_n$ is the electron mobility in the $n$--channel and $C_{ox}$ is the capacitance of the silicon dioxide dielectric \cite{sedra}. For low values of $V_{DS}$, the relationship between $I_{DS}$ and $V_{DS}$ in the triode region is approximately linear. When $V_{DS}$ exceeds $V_{OV}$, the channel pinchoff occurs, and the MOSFET enters the saturation region, in which the current $I_{DS}$ takes the form

\begin{equation}
I_{DS} = \frac{1}{2} \mu_n C_{ox} \frac{W}{L} V_{OV}^2.
\label{eq:sat}
\end{equation} 

Due to channel pinchoff, the drain-source current no longer depends on the voltage $V_{DS}$, and is said to be ``saturated". The full characteristic $I_{DS}$--$V_{DS}$ dependence of an ideal MOSFET when it is turned on ($V_{GS} > V_{TH}$) is illustrated in Fig. 21. For digital applications, where the MOSFET is used as a switch to realize logic gates, the cutoff and triode regions of the FET are utilized. On the other hand, for analog applications, where the MOSFET is used as an amplifier, the saturation region of the FET is utilized \cite{sedra}. 

\begin{figure}[H]
\begin{center}
\includegraphics[scale=0.4]{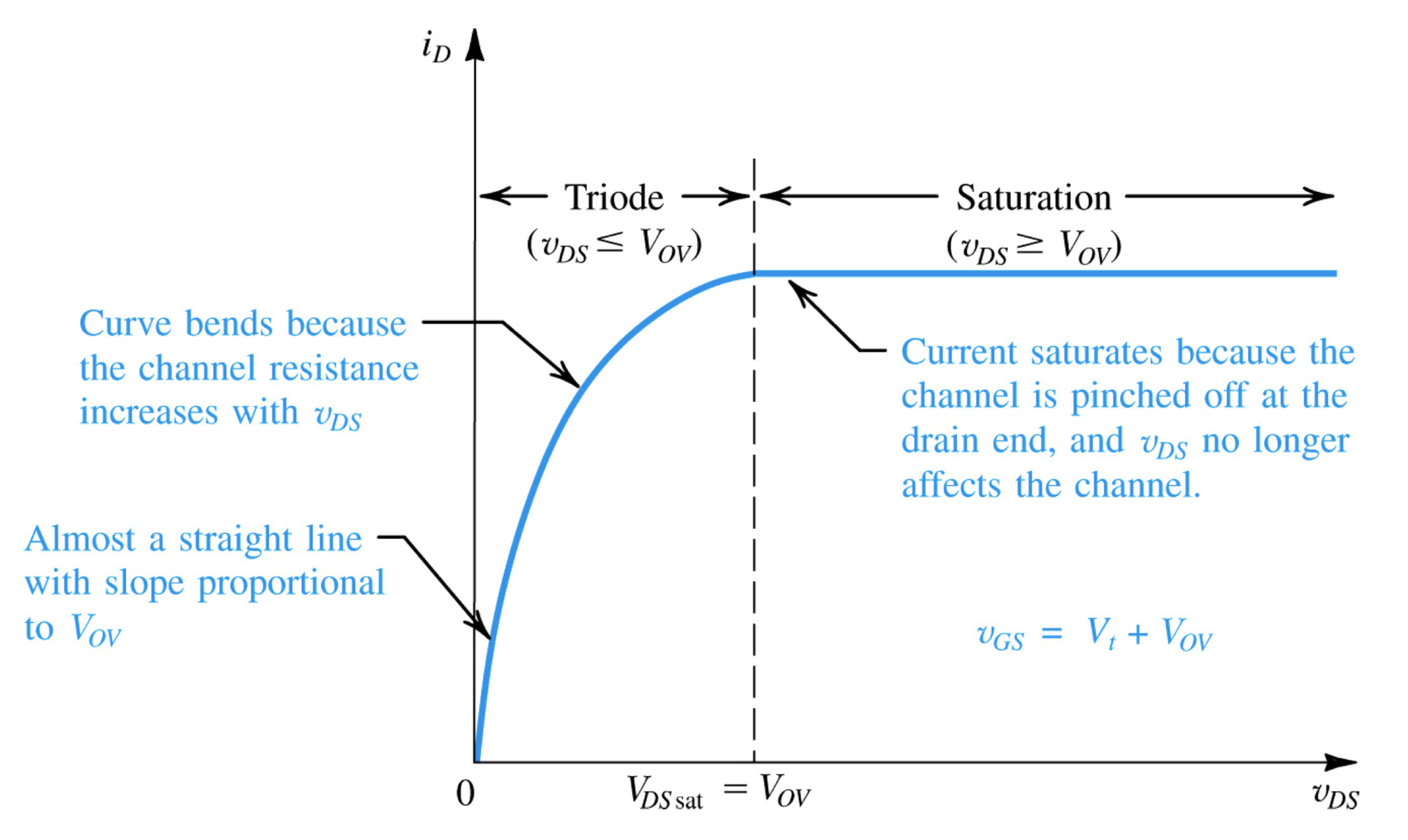}
\caption{A plot of the drain current as a function of the drain--source voltage for an ideal $n$--channel MOSFET. In the triode region, the current varies according to Eq. (18). In the saturation region, the current is constant with respect to $V_{DS}$, and is given by Eq. (19) \cite{sedra}.}
\label{fig:moore}
\end{center}
\end{figure}

Two important FET figures of merit that characterize the speed of a FET in high frequency analog applications (such as terahertz detectors) are the cutoff frequency, $f_c$, and the maximum frequency of oscillation, $f_{osc}$, given by 

\begin{equation}
f_c = \frac{g_m}{2\pi (C_{GS} + C_{GD})}
\label{eq:cut}
\end{equation} 
and 
\begin{equation}
f_{osc} =  \frac{g_m}{4\pi C_{GS} \sqrt{g_{DS} R}}
\label{eq:osc}
\end{equation} 
respectively, where $g_m = \partial I_{DS} / \partial V_{GS}$ is the transconductance parameter, $g_{DS} = \partial I_{DS} / \partial V_{GS}$ is the channel conductance, $C_{GD}$ is the capacitance between the gate and drain terminals, $C_{GS}$ is the capacitance between the gate and source terminals, and $R$ is the gate charging resistance induced by the dielectric \cite{sedra, review, dissertation}. It should be noted that the transconductance parameter, $g_m$, is proportional to the mobility of the $n$--channel, $\mu_n$, and inversely proportional to the channel length, $L$. Thus, both $f_c$ and $f_{osc}$ are proportional to $\mu_n / L$. 

For digital applications where FETs are used to realize logic gates, an important figure of merit that measures the performance of a MOSFET is the on--to--off current ratio, which shall be denoted by $\lambda$ in this report. A large value of $\lambda$ indicates high performance and low power leakage. Low power leakage is a highly desirable property for a FET to have; for example, in portable electronics where an importance is placed on the battery life of a device \cite{sedra, review, dissertation, review1}. 

\pagebreak

\section{State-of-the-Art Graphene FETs}
There is an urgent need for post-silicon technology in industry given the saturation of Moore's law, and incorporating graphene based materials into existing CMOS technology is believed to be a potential solution. Moreover, as stated in the introduction, one of the modern challenges of RF engineering is designing modulators and detectors that work at the untapped terahertz gap (frequencies ranging from $\SI{0.1}{T Hz}$ to $\SI{10} {THz}$). Although mobilities of other novel devices are on the order of $\SI{1e4}{\cm\squared \per \V \per \s}$, which is higher than that of conventional CMOS devices made of silicon, they are  currently not suitable for untapped terahertz applications due to their high cost. As discussed in section 2, graphene exhibits very high mobilities that can reach up to $\SI{2e5}{\cm\squared \per \V \per \s}$ in ideal samples, making it a suitable candidate for use in FETs that are required for high frequency electronics \cite{dissertation}.

One of the figures of merit introduced in section 2.1 is the on--to--off current ratio, $\lambda$. Modern digital electronics applications require a value of $\lambda$ on the order of $10^3$ to $10^4$ \cite{lambda}. A large emphasis was placed on energy bandgaps of graphene and related structures in section 2. This is because a nonzero energy bandgap is essential for digital electronics applications, and a large energy bandgap corresponds to a large value of $\gamma$ \cite{review1, review, dissertation}. This rules out the use of monolayer graphene for digital applications. It is, however, suitable in the realm of high frequency electronics, for which a large value of $\lambda$ is not a requirement \cite{blg, mlgfet}. 

In broad terms, graphene FETs can be classified into two families \cite{dissertation}. The first class of graphene FET implementations involves the use of graphene as a FET channel for carrying current. This class of graphene FETs is typically implemented in one of three different configurations; namely the back-gated, top-gated, and dual-gated configurations \cite{gates}, as illustrated in Fig. 22. In each of these configurations, graphene is used to form the current-carrying channel between the source and the drain. In back-gated and dual-gated graphene FET configurations, a highly doped $\mathrm{Si}$ substrate is used. In back-gated graphene FETs, the substrate acts as the back gate of the FET, whereas in dual-gated graphene FETs, a dielectric layer is deposited on top of the graphene channel, forming a top gate in addition to the back gate. In top-gated graphene FETs, graphene is grown epitaxially on a $\mathrm{SiC}$ substrate, and a dielectric is deposited on top of the graphene channel to form the top gate of the device. Some of the dielectrics used include $\mathrm{SiO_2}$, $\mathrm{Al_2O_3}$, and $\mathrm{HfO_2}$. 

\begin{figure}[H]
\begin{center}
\includegraphics[scale=0.55]{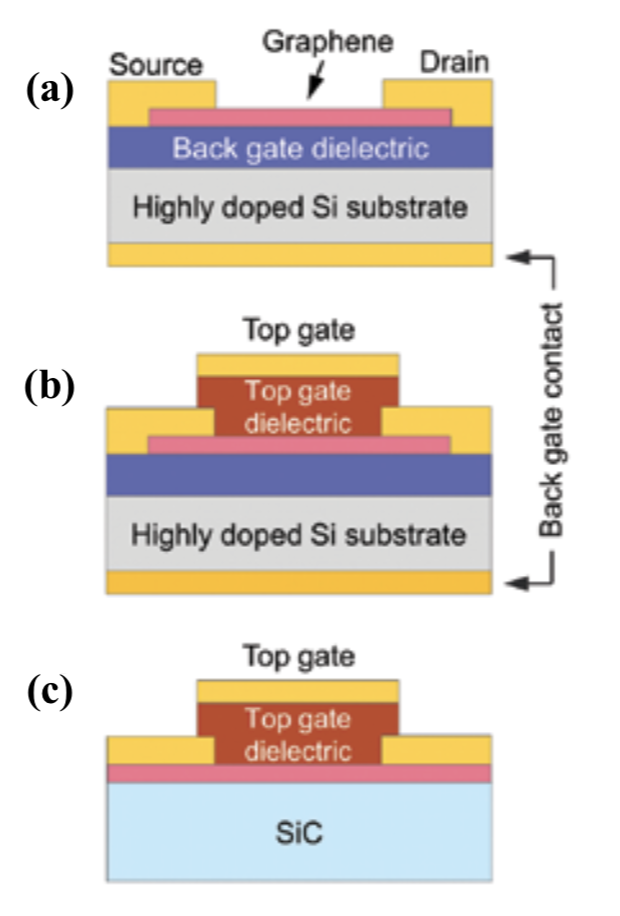}
\caption{Illustrations of the cross sections of (a) a bottom-gated graphene FET, (b) a dual-gated graphene FET, and (c) A top-gated graphene FET \cite{gates}.}
\label{fig:moore}
\end{center}
\end{figure}
Another class of graphene FETs, which is not discussed in this report, hinges on the phenomenon of quantum tunneling. This section only focuses on FETs with monolayer graphene, bilayer graphene, and GNR channels. More information on tunneling graphene FET implementations as well as FETs with other carbon-based channels (such as CNTs, graphene oxide, and graphene nanomeshes) can be found in \cite{review}. 

\subsection{Monolayer Graphene FETs}
The monolayer graphene FET was first demonstrated and studied by Lemme \etal \, in 2007 \cite{lemme}, three years after the discovery of graphene and its ambipolar behavior. One of the key applications of monolayer graphene FETs is high frequency electronics, particularly in the untapped terahertz gap \cite{thz1, thz2, thzreview}.  In fact, as stated in section 3, the figures of merit $f_{c}$ and $f_{osc}$ which determine the speed of a FET in high frequency applications are in fact proportional to the carrier mobility in the FET channel. As such, the parameters $f_c$ and $f_{osc}$ (and, by extension, the mobility, $\mu_n$, and channel length, $L$) introduced in section 3.1 are of key interest in this context. One of the challenges, however, is that although monolayer graphene exhibits high mobility, its mobility is degraded by the dielectric and substrates used, in addition to degradation that results from the synthesis techniques outlined in section 2.3. A short channel length is therefore desirable for maximizing values of $f_c$ and $f_{osc}$. A cross-sectional schematic of a monolayer graphene FET is shown in Fig. 23. 

\begin{figure}[H]
\begin{center}
\includegraphics[scale=0.4]{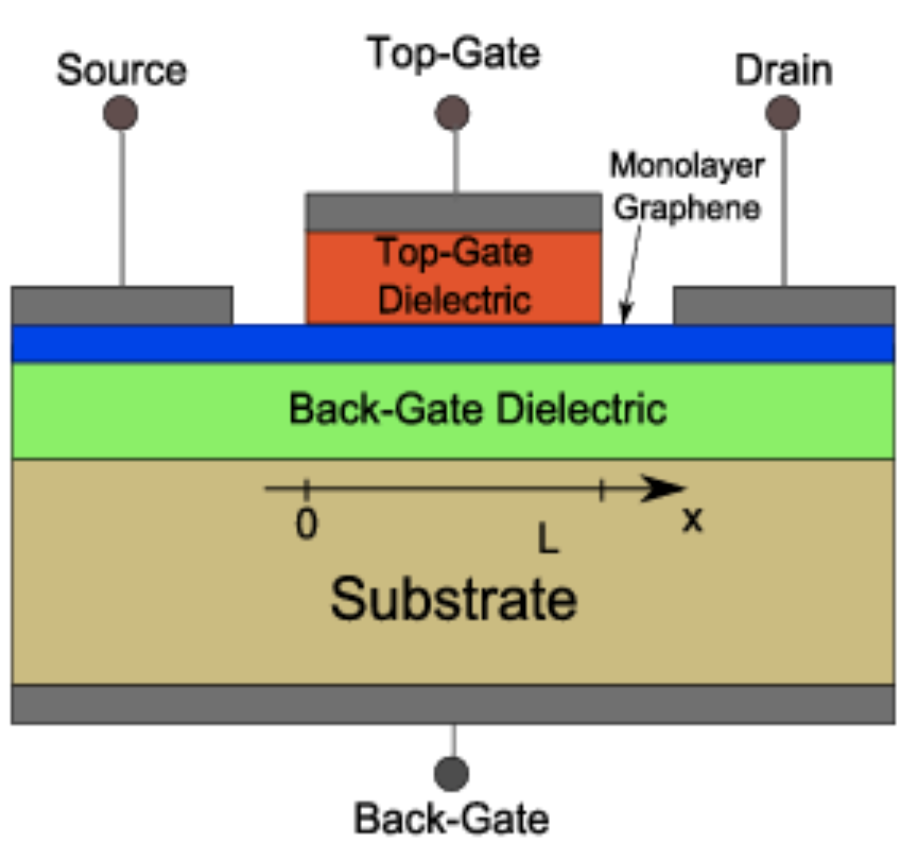}
\caption{A cross-sectional diagram of a monolayer graphene FET in the dual-gated configuration \cite{mlgfet}.}
\label{fig:moore}
\end{center}
\end{figure}

Meric \etal \, reported the first instance of a high frequency measurement of a monolayer graphene FET in 2008, with $f_c = \SI{14.7}{GHz}$ and $L = \SI{500}{nm}$ \cite{meric}. Two years later, monolayer graphene FETs with $f_c = \SI{100}{GHz}$ and $L = \SI{240}{nm}$ \cite{100} as well as $f_c = \SI{300}{GHz}$ and $L = \SI{144}{nm}$ \cite{nanogate} were realized, the latter using a nanowire gate in order to retain a large value of mobility. In 2012, a monolayer graphene FET with a nanowire gate was demonstrated by Cheng \etal \, with $f_c = \SI{427}{GHz}$, which is the highest achieved value of $f_c$ to date, and $L = \SI{67}{nm}$ \cite{427, dissertation}. This value of $f_c$, which is currently the state-of-the-art for graphene FETs, is comparable with that of $\mathrm{InP}$ and $\mathrm{GaAs}$ high electron mobility transistors (HEMTs) \cite{model, comparable}. In the past few years, advancements have been made in using monolayer graphene FETs to realize high frequency electronics. For example, in 2017, a $\SI{400}{GHz}$ monolayer graphene FET detector with high responsitivity was realized \cite{thz2}. In the same year, Yang \etal \, demonstrated a monolayer graphene FET detector capable of terahertz detection at room temperature from $\SI{330}{GHz}$ to $\SI{500}{GHz}$ \cite{thzdetect}. In 2018, graphene FETs and plasmons were used for resonant terahertz radiation detection \cite{plasm}.

Progress in increasing $f_{osc}$ in monolayer graphene FETs has been slower; values of $f_{osc}$ for monolayer graphene FETs typically range from $\SI{30}{GHz}$ to $\SI{200}{GHz}$, showing poorer performance than conventional $\mathrm{Si}$-based FETs \cite{dissertation}. This is a result of the fact that, as can be seen from Eq. (\ref{eq:osc}), a large value of $f_{osc}$ requires a small value of $g_{DS}$, the channel conductance. The model of a conventional MOSFET such as that presented in section 2.1 displays a $I_{DS}$--$V_{DS}$ shown in Fig. 21, where the current enters a saturation region when $V_{DS} > V_{OV}$. However, graphene FETs display a more peculiar characteristic, in which increasing $V_{DS}$ beyond a certain value causes $I_{DS}$ to vary linearly exit the saturation region, increasing the value of $g_{DS}$, which leads to a smaller value of $f_{osc}$ \cite{review, dissertation, meric}. This is a result of interband tunneling and the quasi-ballistic nature of carrier transport within graphene \cite{blg}. There are several engineering research groups that have studied and modeled the effects of non-ideal $I_{DS}$--$V_{DS}$ characteristics and other phenomena such as negative differential resistance in monolayer graphene FETs \cite{ndr, compact}. 

\subsection{Bilayer Graphene FETs}
Another way to implement a graphene FET is to use a bilayer graphene channel.  A cross-sectional schematic of a bilayer graphene FET is shown in Fig. 24.  

\begin{figure}[H]
\begin{center}
\includegraphics[scale=0.45]{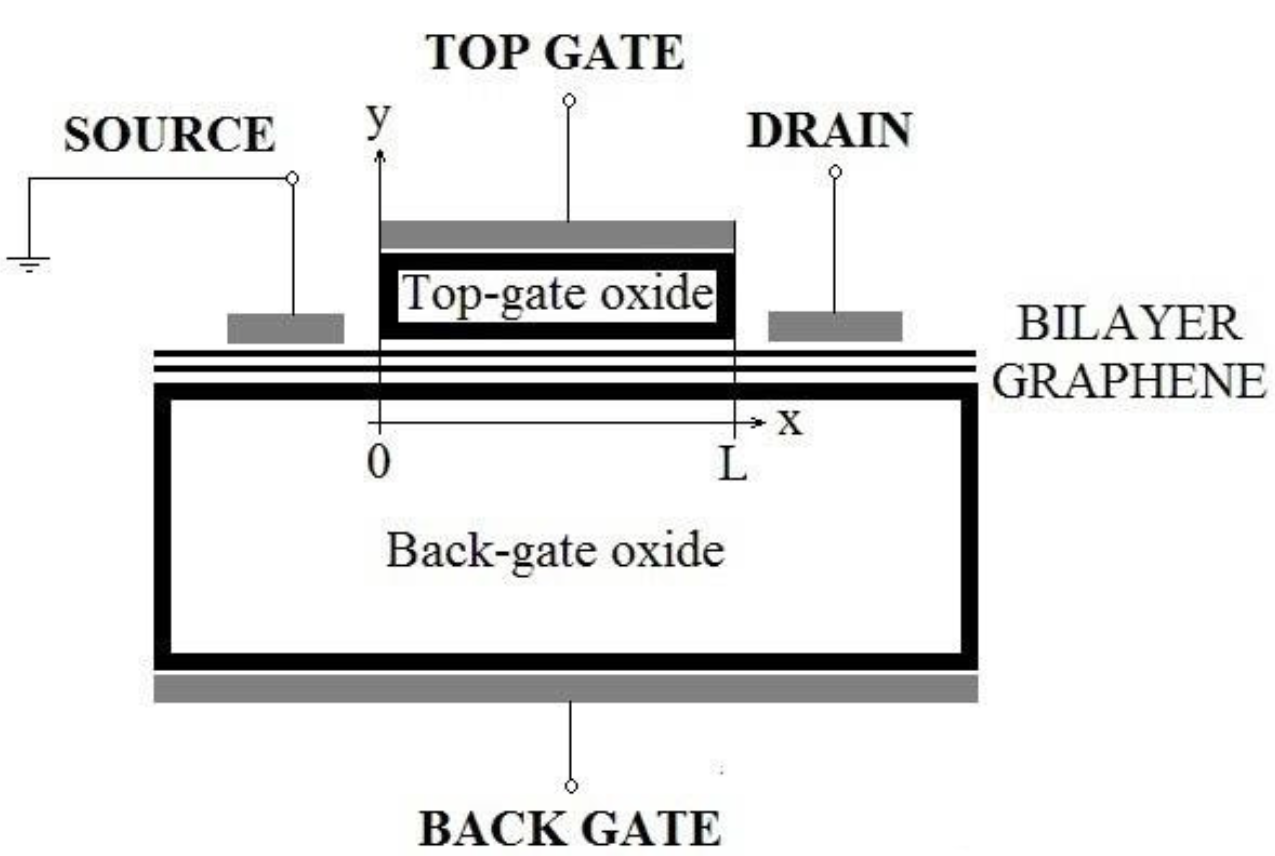}
\caption{A cross-sectional diagram of a bilayer graphene FET in the dual-gated configuration \cite{formula}.}
\label{fig:moore}
\end{center}
\end{figure}

Although bilayer graphene typically exhibits a lower mobility than monolayer graphene, the use of a bilayer graphene channel in FETs offers some advantages over monolayer graphene. In particular, bilayer graphene FETs have been shown to possess a larger intrinsic voltage gain than monolayer graphene FETs \cite{dissertation}. Moreover, the bandgap induced in bilayer graphene by applying a perpendicular electric field has been shown to improve current saturation and the maximum frequency of oscillation, $f_{osc}$ \cite{blg, volt}. This is because the existence of a nonzero bandgap in bilayer graphene (upon the application of a perpendicular electric field) suppresses interband tunneling. Furthermore, bilayer graphene FETs show a leakage current that is orders of magnitude lower than that of a typical monolayer graphene FET at low temperatures \cite{blg3}. Although the gap  in leakage currents between the two FET devices decreases at higher temperatures, a lower leakage current is desirable in both analog and digital applications. 

The zero bandgap of monolayer graphene implies a small value of $\lambda$ ($\approx 5$ for top-gated FETs) which is unsuitable for digital applications \cite{dissertation}. As stated in section 2, bandgaps as large as $\SI{130}{meV}$ have been demonstrated in bilayer graphene. For bilayer graphene FETs, this corresponds to a value of $\lambda \approx 10^2$ \cite{comparable}. While this is  an improvement over the values of $\lambda$ observed in monolayer graphene FETs, it is not sufficient for modern applications in digital electronics, which require a minimum value of $\lambda$ on the order of $10^3$ to $10^4$. 

\subsection{Graphene Nanoribbon FETs}
An alternative to using bilayer graphene as a means of achieving a larger value of $\lambda$ is to use GNR FETs. A GNR FET has a similar structure to monolayer graphene and bilayer graphene FETs; an armchair GNR used as a current carrying channel in the FET device, as depicted in Fig. 25.  

\begin{figure}[H]
\begin{center}
\includegraphics[scale=1]{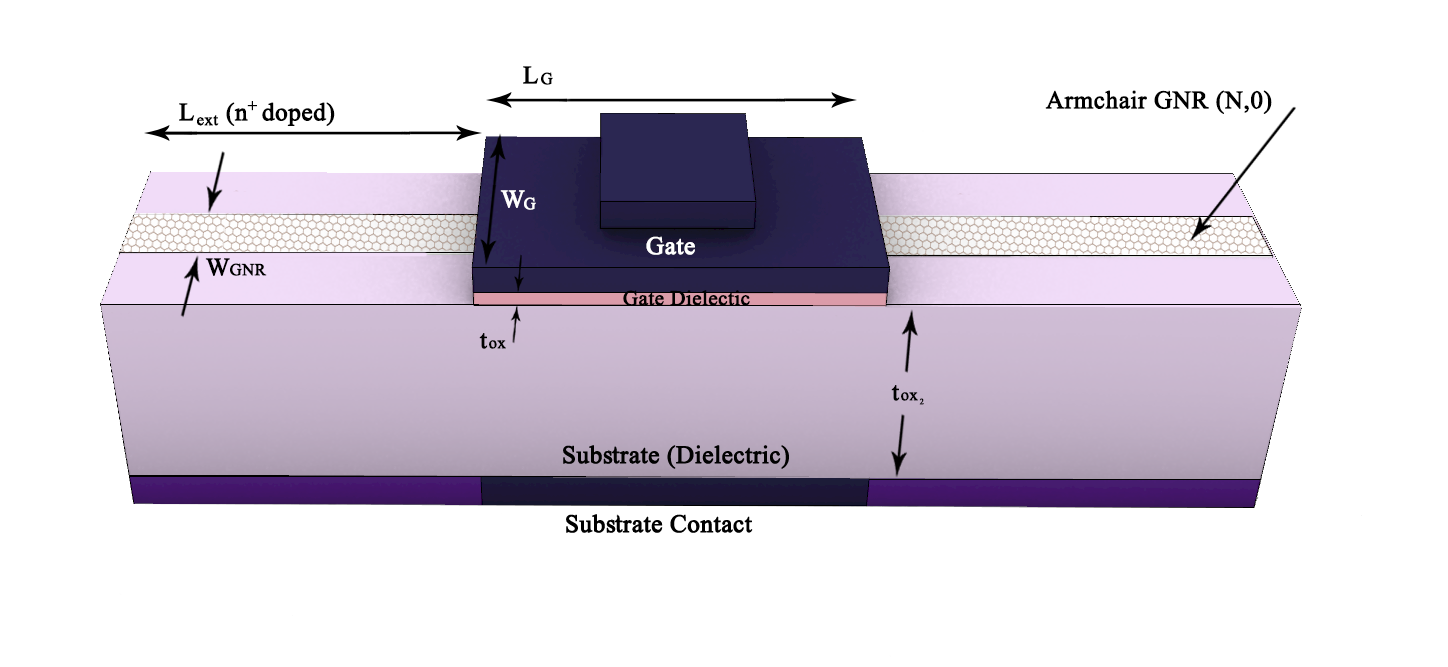}
\caption{A diagram of a GNR FET with a single channel \cite{gnrdiag}.}
\label{fig:moore}
\end{center}
\end{figure}

Bandgaps as large as $\SI{2.3}{eV}$ have been observed in armchair graphene nanoribbons \cite{2.3}, which is approximately three orders of magnitude larger than the largest bandgaps observed in bilayer graphene under the application of a perpendicular electric field \cite{dissertation}. In fact, values of $\lambda$ as high as $10^7$ have been demonstrated in sub-$\SI{10}{nm}$ width $p$--type GNR FETs \cite{e7} -- outperforming bilayer graphene FETs by five orders of magnitude. Another advantage of GNRs is that their small transverse width allows multiple GNRs to be used as channels on a single device. This has the benefit of increasing the drive current and enhancing switching characteristics for high performance applications \cite{jc}. 

Since GNR fabrication technology is still in its infancy, much of the performance issues observed in GNR FETs are limited by non-idealities in GNR samples. In particular, fabricating well-defined GNRs with high precision is not an easy task, and the existence of zigzag edges in armchair GNR samples, can degrade the performance of a GNR FET \cite{jc}. Furthermore, although $p$--type GNR FETs with large values of $\lambda$ have been demonstrated, digital applications also require high performance $n$--type GNR FETs \cite{dissertation}. Furthermore, mobility degradation is one of the biggest disadvantages of GNR FETs -- for large values of $\lambda$ in the range from $10^4$ to $10^7$, GNRs must possess sub-$\SI{10}{nm}$ width, which results in carrier mobilities lower than $\SI{1e3}{\cm\squared \per \V \per \s}$ due to phonon scattering near the edges of the GNR.
\pagebreak

\section{Conclusion and Future Perspectives}
Although graphene exhibits remarkable electronic properties that make it a suitable candidate for replacing silicon and extending the lifetime of Moore's law, there remains a lot of research to be conducted around overcoming the challenges associated with realizing graphene FETs in industry. Among the challenging aspects of implementing graphene FET technology on a large scale is the trade-off between scalability and quality of graphene samples associated with different synthesis techniques. As discussed in section 2, CVD is the most scalable and least costly technique for synthesizing graphene layers in industry, but results in samples with relatively low mobilities; making it difficult to harness the potential of graphene as a high-mobility alternative to silicon. 

Another key trade-off that manifests itself in this research area is that of bandgap engineering and how opening a bandgap in graphene by using bilayer graphene or GNRs, as discussed in sections 2 and 4, inevitably results in FETs with much lower mobilities than monolayer graphene FETs. The zero bandgap of graphene is problematic for electronic applications. Evidently, bandgap engineering is crucial for digital electronics, and of the implementations presented in this report, GNR FETs show the most promise toward that end, with observed on--to--off current ratios reaching $10^7$, although there remains a lot of work to be done in enhancing the fabrication processes by which GNRs are made, and overcoming mobility degradation in GNR samples. 

It is evident, based on the state-of-the-art review presented in section 4, that the real potential of graphene FETs in the near future lies in high frequency applications. The highest observed value of $f_c$ to date in monolayer graphene FETs is $\SI{427}{GHz}$, which is comparable to that of alternative post-silicon technologies such as $\mathrm{InP}$ and $\mathrm{GaAs}$ HEMTs, and superior to existing conventional CMOS technologies. Moreover, terahertz graphene FET detectors, operating at frequencies ranging from $\SI{300}{GHz}$ to $\SI{400}{GHz}$ have been demonstrated, which is very promising and indicative of the prospects of using graphene FETs for terahertz detectors in the near future.

Although this report examined a few examples of graphene FET implementations, it is important to note that researchers have been exploring a much wider variety of graphene (or carbon-based) FET implementations, such as carbon nanotube FETs, graphene oxide FETs, graphene nanomeshes, and vertical tunneling FETs. In fact, graphene is no longer the only two dimensional material of interest to scientists and engineers. More recently, researchers have been examining other novel two dimensional structures such as graphyne and silicene, which may offer advantages over graphene in terms of bandgap engineering \cite{review}. Overall, at present, it is unclear whether graphene will ever replace silicon in modern consumer electronics at large, for the aforementioned reasons regarding the difficulty of bandgap engineering and the synthesis of high mobility graphene samples on a large scale. Nevertheless, it is becoming more apparent that graphene could play an important role in more specialized areas of modern electronic engineering, such as terahertz technology.

\pagebreak

\bibliographystyle{ieee}

\bibliography{Review}

\pagebreak

\section*{Appendix}
\addcontentsline{toc}{section}{Appendix}
\renewcommand{\theequation}{A.\arabic{equation}}
\fancyfoot[CO,CE]{\thepage}
\setcounter{equation}{0}
\pagestyle{abstract}
In this appendix, the electronic band structure of graphene is derived using the tight binding method. This derivation has largely been adapted from \cite{notes}.

As described in section 2, a unit cell in graphene contains two atoms -- each from one of the interpenetrating sublattices. Suppose that the two sublattices are labeled by $A$ and $B$, in accordance with Fig. 4. Then, the Bloch functions associated with sublattices $A$ and $B$ may be defined by

\begin{equation}
\psi_{\mathbf{k}}^{(A)} (\mathbf{r}) = \frac{1}{\sqrt{N}} \sum_{\mathbf{R}_{A}}^{} e^{i \mathbf{k} \cdot \mathbf{R}_{A}} \phi_{A} (\mathbf{r} - \mathbf{R}_A)
\end{equation}
and
\begin{equation}
\psi_{\mathbf{k}}^{(B)} (\mathbf{r}) = \frac{1}{\sqrt{N}} \sum_{\mathbf{R}_{B}}^{} e^{i \mathbf{k} \cdot \mathbf{R}_{B}} \phi_{B} (\mathbf{r} - \mathbf{R}_B)
\end{equation}
respectively, where $N$ is the number of unit cells, $\mathbf{r}$ is a position vector, $\mathbf{k}$ is the wave vector associated with crystal momentum $\hbar \mathbf{k}$ , $\mathbf{R}_{\alpha}$ is the lattice vector associated with a carbon atom in the sublattice $\alpha$, and $\phi_{\alpha} (\mathbf{r} - \mathbf{R}_{\alpha})$ is a normalized eigenstate of the Hamiltonian of a carbon atom in the sublattice $\alpha$. The trial wave function, $\psi_{\mathbf{k}}$, may then be written as 

\begin{equation}
\psi_{\mathbf{k}} (\mathbf{r}) = a_{\mathbf{k}} \psi_{\mathbf{k}}^{(A)} (\mathbf{r}) + b_{\mathbf{k}} \psi_{\mathbf{k}}^{(A)} (\mathbf{r}),
\end{equation}
where $a_{\mathbf{k}}, b_{\mathbf{k}} \in \mathbb{C}$ are complex coefficients that depend on $\mathbf{k}$. Substituting the trial wave function into the time-independent Schrödinger equation, one obtains

\begin{equation}
\hat{H} \psi_{\mathbf{k}} = \epsilon_{\mathbf{k}} \psi_{\mathbf{k}},
\end{equation}
where $\hat{H}$ is the Hamiltonian operator and $\epsilon_{\mathbf{k}}$ is the energy eigenvalue associated with $\mathbf{k}$. Multiplying both sides of Eq. (A.4) by $\psi_{\mathbf{k}}^{\dagger}$ and carrying out a spatial integration over the unit cell yields

\begin{equation}
\begin{pmatrix} a_{\mathbf{k}}^* & b_{\mathbf{k}}^* \end{pmatrix} \mathcal{H}_{\mathbf{k}} \begin{pmatrix} a_{\mathbf{k}}^* \\ b_{\mathbf{k}}^* \end{pmatrix} = \epsilon_{\mathbf{k}} \begin{pmatrix} a_{\mathbf{k}}^* & b_{\mathbf{k}}^* \end{pmatrix} \mathcal{S}_{\mathbf{k}} \begin{pmatrix} a_{\mathbf{k}}^* \\ b_{\mathbf{k}}^* \end{pmatrix},
\end{equation}
where $\mathcal{H}_{\mathbf{k}}$ is the matrix representation of the Hamiltonian operator in the $\left(\vert \psi_{\mathbf{k}}^{(A)} \rangle, \vert \psi_{\mathbf{k}}^{(B)} \rangle\right)$ basis, namely,

\begin{equation}
\mathcal{H}_{\mathbf{k}} = \begin{pmatrix}
\langle \psi_{\mathbf{k}}^{(A)} \vert \hat{H} \vert \psi_{\mathbf{k}}^{(A)} \rangle & \langle \psi_{\mathbf{k}}^{(A)} \vert \hat{H} \vert \psi_{\mathbf{k}}^{(B)} \rangle  \\
\langle \psi_{\mathbf{k}}^{(B)} \vert \hat{H} \vert \psi_{\mathbf{k}}^{(A)} \rangle & \langle \psi_{\mathbf{k}}^{(B)} \vert \hat{H} \vert \psi_{\mathbf{k}}^{(B)} \rangle
\end{pmatrix},
\end{equation}
and $\mathcal{S}_{\mathbf{k}}$ is the overlap matrix (which takes into account the fact that $\left(\vert \psi_{\mathbf{k}}^{(A)} \rangle, \vert \psi_{\mathbf{k}}^{(B)} \rangle\right)$ is non-orthogonal), given by

\begin{equation}
\mathcal{S}_{\mathbf{k}} = \begin{pmatrix}
\langle \psi_{\mathbf{k}}^{(A)} \vert \psi_{\mathbf{k}}^{(A)} \rangle & \langle \psi_{\mathbf{k}}^{(A)} \vert \psi_{\mathbf{k}}^{(B)} \rangle  \\
\langle \psi_{\mathbf{k}}^{(B)} \vert \psi_{\mathbf{k}}^{(A)} \rangle & a\langle \psi_{\mathbf{k}}^{(B)} \vert \psi_{\mathbf{k}}^{(B)} \rangle
\end{pmatrix}.
\end{equation}
The energy dispersion relation $\epsilon (\mathbf{k}) = \epsilon_{\mathbf{k}}$ may then be determined by solving the secular equation
\begin{equation}
\mathrm{det} \left( \mathcal{H}_{\mathbf{k}} - \epsilon_{\mathbf{k}} \mathcal{S}_{\mathbf{k}} \right) = 0,
\end{equation}
which holds for nonzero, physically admissable trial wave functions $\psi_{\mathbf{k}}$ (\ie, $a_{\mathbf{k}}, b_{\mathbf{k}} \neq 0$). 

The diagonal elements of $\mathcal{H}_{\mathbf{k}}$ are given by

\begin{equation}
\left( \mathcal{H}_{\mathbf{k}} \right)_{AA} = \frac{1}{N} \mathlarger{\sum}_{\mathbf{R}_A} \mathlarger{\sum}_{\mathbf{R}_A'} e^{i \mathbf{k} \cdot  (\mathbf{R}_A' - \mathbf{R}_A)} \langle \phi_A (\mathbf{r} - \mathbf{R}_A) \vert \hat{H} \vert \phi_A (\mathbf{r} - \mathbf{R}_A') \rangle
\end{equation}
and 
\begin{equation}
\left( \mathcal{H}_{\mathbf{k}} \right)_{BB} = \frac{1}{N} \mathlarger{\sum}_{\mathbf{R}_B} \mathlarger{\sum}_{\mathbf{R}_B'} e^{i \mathbf{k} \cdot  (\mathbf{R}_B' - \mathbf{R}_B)} \langle \phi_B (\mathbf{r} - \mathbf{R}_B) \vert \hat{H} \vert \phi_B (\mathbf{r} - \mathbf{R}_B') \rangle.
\end{equation}
By assuming that the only contribution comes from the same unit cell; \ie, $\mathbf{R}_A' = \mathbf{R}_A$ and $\mathbf{R}_B' = \mathbf{R}_B$, the diagonal elements approximately reduce to 

\begin{equation}
\left( \mathcal{H}_{\mathbf{k}} \right)_{AA} \approx \frac{1}{N} \mathlarger{\sum}_{\mathbf{R}_A} \langle \phi_A (\mathbf{r} - \mathbf{R}_A) \vert \hat{H} \vert \phi_A (\mathbf{r} - \mathbf{R}_A) \rangle
\end{equation}
and

\begin{equation}
\left( \mathcal{H}_{\mathbf{k}} \right)_{BB} \approx \frac{1}{N} \mathlarger{\sum}_{\mathbf{R}_B} \langle \phi_B (\mathbf{r} - \mathbf{R}_B) \vert \hat{H} \vert \phi_B (\mathbf{r} - \mathbf{R}_B) \rangle.
\end{equation}
The terms in the summations in Eq. (A.11) and Eq. (A.12) are constant, and the atoms on sublattices $A$ and $B$ are chemically indistinguishable, so one may write

\begin{equation}
\epsilon_0 = \langle \phi_A (\mathbf{r} - \mathbf{R}_A) \vert \hat{H} \vert \phi_A (\mathbf{r} - \mathbf{R}_A) \rangle = \langle \phi_B (\mathbf{r} - \mathbf{R}_B) \vert \hat{H} \vert \phi_B (\mathbf{r} - \mathbf{R}_B) \rangle,
\end{equation}
where $\epsilon_0$ is an energy parameter associated with the atomic eigenstates $\phi_A$ and $\phi_B$. It follows that the diagonal elements of the Hamiltonian matrix reduce to 

\begin{equation}
\left( \mathcal{H}_{\mathbf{k}} \right)_{AA} = \left( \mathcal{H}_{\mathbf{k}} \right)_{BB} = \epsilon_0,
\end{equation}
by noting that the sums in Eq. (A.11) and Eq. (A.12) are sums over $N$ unit cells. The diagonal elements of $\mathcal{S}_{\mathbf{k}}$ are given by

\begin{equation}
\left( \mathcal{S}_{\mathbf{k}} \right)_{AA} = \frac{1}{N} \mathlarger{\sum}_{\mathbf{R}_A} \mathlarger{\sum}_{\mathbf{R}_A'} e^{i \mathbf{k} \cdot  (\mathbf{R}_A' - \mathbf{R}_A)} \langle \phi_A (\mathbf{r} - \mathbf{R}_A) \vert  \phi_A (\mathbf{r} - \mathbf{R}_A') \rangle
\end{equation}
and 

\begin{equation}
\left( \mathcal{S}_{\mathbf{k}} \right)_{BB} = \frac{1}{N} \mathlarger{\sum}_{\mathbf{R}_B} \mathlarger{\sum}_{\mathbf{R}_B'} e^{i \mathbf{k} \cdot  (\mathbf{R}_B' - \mathbf{R}_B)} \langle \phi_B (\mathbf{r} - \mathbf{R}_B) \vert  \phi_B (\mathbf{r} - \mathbf{R}_B') \rangle.
\end{equation}
Applying the same approximation (assuming that the dominant contribution comes from the same unit cell), the diagonal terms approximately reduce to

\begin{equation}
\left( \mathcal{S}_{\mathbf{k}} \right)_{AA} \approx \frac{1}{N} \mathlarger{\sum}_{\mathbf{R}_A}\langle \phi_A (\mathbf{r} - \mathbf{R}_A) \vert  \phi_A (\mathbf{r} - \mathbf{R}_A) \rangle
\end{equation}
and

\begin{equation}
\left( \mathcal{S}_{\mathbf{k}} \right)_{BB} \approx \frac{1}{N} \mathlarger{\sum}_{\mathbf{R}_B}\langle \phi_B (\mathbf{r} - \mathbf{R}_B) \vert  \phi_B (\mathbf{r} - \mathbf{R}_B) \rangle.
\end{equation}
Since the eigenstates $\phi_A$ and $\phi_B$ are normalized, 

\begin{equation}
\langle \phi_A (\mathbf{r} - \mathbf{R}_A) \vert  \phi_A (\mathbf{r} - \mathbf{R}_A) \rangle = \langle \phi_B (\mathbf{r} - \mathbf{R}_B) \vert  \phi_B (\mathbf{r} - \mathbf{R}_B) \rangle = 1,
\end{equation}
and, by noting that the summations in Eq. (A.17) and Eq. (A.18) are over $N$ unit cells,

\begin{equation}
\left( \mathcal{S}_{\mathbf{k}} \right)_{AA}  = \left( \mathcal{S}_{\mathbf{k}} \right)_{BB}  = 1.
\end{equation}

Figure 4 shows the how the position a carbon atom in the $A$ sublattice relative to its three neighboring carbon atoms in the $B$ sublattice. In particular, with respect to the coordinate system $(\mathbf{\hat{e}}_1, \mathbf{\hat{e}}_2, \mathbf{\hat{e}}_3)$, the three displacement vectors $\mathbf{v}_1$, $\mathbf{v}_2$,and $\mathbf{v}_3$ from an $A$ atom to its three nearest neighbors ($B$ atoms) may be written as

\begin{equation}
\mathbf{v}_1 = a \mathbf{\hat{e}}_2,
\end{equation}

\begin{equation}
\mathbf{v}_2 = \frac{1}{2}(a' \mathbf{\hat{e}}_1 - a \mathbf{\hat{e}}_2),
\end{equation}
and

\begin{equation}
\mathbf{v}_3 = -\frac{1}{2}(a' \mathbf{\hat{e}}_1 + a \mathbf{\hat{e}}_2),
\end{equation}
where $a \approx \SI{1.42}{\angstrom}$ is the interatomic spacing between the carbon atoms and $a' = a \sqrt{3}$ is the lattice constant. Considering the off-diagonal matrix element $\left( \mathcal{H}_{\mathbf{k}} \right)_{AB}$

\begin{equation}
\left( \mathcal{H}_{\mathbf{k}} \right)_{AB} = \frac{1}{N} \mathlarger{\sum}_{\mathbf{R}_A} \mathlarger{\sum}_{\mathbf{R}_B} e^{i \mathbf{k} \cdot  (\mathbf{R}_B - \mathbf{R}_A)} \langle \phi_A (\mathbf{r} - \mathbf{R}_A) \vert \hat{H} \vert \phi_A (\mathbf{r} - \mathbf{R}_B) \rangle,
\end{equation}
the inner sum may be approximated by 

\begin{equation}
\mathlarger{\sum}_{\mathbf{R}_B} e^{i \mathbf{k} \cdot  (\mathbf{R}_B - \mathbf{R}_A)} \langle \phi_A (\mathbf{r} - \mathbf{R}_A) \vert \hat{H} \vert \phi_A (\mathbf{r} - \mathbf{R}_B) \rangle \approx  \mathlarger{\sum}_{j = 1}^3 e^{i \mathbf{k} \cdot \mathbf{v}_j } \langle \phi_A (\mathbf{r} - \mathbf{R}_A) \vert \hat{H} \vert \phi_A (\mathbf{r} - (\mathbf{R}_A + \mathbf{v}_j)) \rangle
\end{equation}
if the only interactions considered are the interactions between the $A$ atom and its three nearest neighbors. Therefore, Eq. (A.24) may be rewritten as 

\begin{equation}
\left( \mathcal{H}_{\mathbf{k}} \right)_{AB} = \frac{1}{N} \mathlarger{\sum}_{j = 1}^3 \mathlarger{\sum}_{\mathbf{R}_A} e^{i \mathbf{k} \cdot \mathbf{v}_j } \langle \phi_A (\mathbf{r} - \mathbf{R}_A) \vert \hat{H} \vert \phi_A (\mathbf{r} - (\mathbf{R}_A + \mathbf{v}_j)) \rangle.
\end{equation}
By noting that $\langle \phi_A (\mathbf{r} - \mathbf{R}_A) \vert \hat{H} \vert \phi_A (\mathbf{r} - (\mathbf{R}_A + \mathbf{v}_j)) \rangle$ is a constant term in the summation, one may write

\begin{equation}
t = \langle \phi_A (\mathbf{r} - \mathbf{R}_A) \vert \hat{H} \vert \phi_A (\mathbf{r} - (\mathbf{R}_A + \mathbf{v}_j)) \rangle,
\end{equation}
where $t  \in \mathbb{R}$ is a tight binding hopping parameter. Therefore, 

\begin{equation}
\left( \mathcal{H}_{\mathbf{k}} \right)_{AB} = t \mathlarger{\sum}_{j = 1}^3  e^{i \mathbf{k} \cdot \mathbf{v}_j } = t \, h(\mathbf{k})
\end{equation}
where $h$ is a complex-valued function of $\mathbf{k}$ defined by

\begin{equation}
h(\mathbf{k}) = \mathlarger{\sum}_{j = 1}^3  e^{i \mathbf{k} \cdot \mathbf{v}_j }.
\end{equation}
The other off-diagonal element, $\left( \mathcal{H}_{\mathbf{k}} \right)_{BA}$, is the complex conjugate of $\left( \mathcal{H}_{\mathbf{k}} \right)_{AB}$. Therefore,

\begin{equation}
\left( \mathcal{H}_{\mathbf{k}} \right)_{BA} = t \, h^*(\mathbf{k}).
\end{equation}

Likewise, the nearest neighbors approximation can be used to show that the off-diagonal elements of $\mathcal{S}_{\mathbf{k}}$ are given by 

\begin{equation}
\left( \mathcal{S}_{\mathbf{k}} \right)_{AB} = s \, h(\mathbf{k})
\end{equation}
and

\begin{equation}
\left( \mathcal{S}_{\mathbf{k}} \right)_{BA} = s \, h^*(\mathbf{k}),
\end{equation}
where  $s  \in \mathbb{R}$ is an overlap parameter given by

\begin{equation}
s = \langle \phi_A (\mathbf{r} - \mathbf{R}_A) \vert \phi_A (\mathbf{r} - (\mathbf{R}_A + \mathbf{v}_j)) \rangle
\end{equation}
for $j \in \set{1,2,3}$. 

By using the results above, the matrices $\mathcal{H}_{\mathbf{k}}$ and $\mathcal{S}_{\mathbf{k}}$ may be written as

\begin{equation}
\mathcal{H}_{\mathbf{k}} = \begin{pmatrix}
\epsilon_0 & t \, h(\mathbf{k})  \\
t \, h^*(\mathbf{k}) & \epsilon_0
\end{pmatrix}
\end{equation}
and

\begin{equation}
\mathcal{S}_{\mathbf{k}} = \begin{pmatrix}
1 & s \, h(\mathbf{k})  \\
s \, h^*(\mathbf{k}) & 1
\end{pmatrix},
\end{equation}
respectively. Substituting $\mathcal{H}_{\mathbf{k}}$ and $\mathcal{S}_{\mathbf{k}}$ into the secular equation (Eq. (A.8)), one obtains the equation

\begin{equation}
(\epsilon_0 - \epsilon_{\mathbf{k}})^2 - (t - s \, \epsilon_{\mathbf{k}})^2 |h(\mathbf{k})|^2 = 0.
\end{equation}
One may define $f(\mathbf{k}) \equiv |h(\mathbf{k})|$. Then, using Eq. (A.29) and Eq. (A.21)--(A.23) $f(\mathbf{k})$ can then be found to be

\begin{equation}
f(\mathbf{k}) = \sqrt{1 + 4 \cos{\left( \frac{3k_y a}{2} \right)} \cos{\left( \frac{\sqrt{3}k_x a}{2} \right)} + 4\cos^2{\left( \frac{\sqrt{3}k_x a}{2} \right)}}.
\end{equation}
Finally, solving for $\epsilon (\mathbf{k}) = \epsilon_{\mathbf{k}}$ in Eq. (A.36) gives the dispersion relation

\begin{equation}
\epsilon^{(\pm)}(\mathbf{k}) = \frac{\epsilon_0 \pm t\, f(\mathbf{k}) }{1 \pm s\, f(\mathbf{k})},
\end{equation}
where $+$ denotes the conduction band while $-$ denotes the valence band.

\end{document}